\documentclass[11pt,a4paper]{article}
\usepackage{jinstpub}	
\pdfoutput=1 	

\usepackage{graphicx} 	
\usepackage{amssymb} 	
\usepackage{mathtools} 	
\usepackage{comment} 	
\usepackage{color} 		
\usepackage{multirow}	
\usepackage{gensymb}
\usepackage{subcaption}
\usepackage{epstopdf}
\usepackage[]{siunitx}  
\DeclareSIUnit\sample{S}
\DeclareSIUnit\bits{bits}

\usepackage{lineno} 	
\newcommand*\patchAmsMathEnvironmentForLineno[1]{%
  \expandafter\let\csname old#1\expandafter\endcsname\csname #1\endcsname
  \expandafter\let\csname oldend#1\expandafter\endcsname\csname end#1\endcsname
  \renewenvironment{#1}%
     {\linenomath\csname old#1\endcsname}%
     {\csname oldend#1\endcsname\endlinenomath}}%
\newcommand*\patchBothAmsMathEnvironmentsForLineno[1]{%
  \patchAmsMathEnvironmentForLineno{#1}%
  \patchAmsMathEnvironmentForLineno{#1*}}%
\AtBeginDocument{%
\patchBothAmsMathEnvironmentsForLineno{equation}%
\patchBothAmsMathEnvironmentsForLineno{align}%
\patchBothAmsMathEnvironmentsForLineno{flalign}%
\patchBothAmsMathEnvironmentsForLineno{alignat}%
\patchBothAmsMathEnvironmentsForLineno{gather}%
\patchBothAmsMathEnvironmentsForLineno{multline}%
}

\usepackage{xspace} 	
\usepackage[all]{hypcap} 

\newcommand{\sep}{, } 

\makeatletter
\renewcommand\tableofcontents{%
    \@starttoc{toc}%
}
\makeatother

\newif\ifshowchanges
\showchangesfalse

\newcommand{\vo}[1]{}

\ifshowchanges
\usepackage{soul}
\soulregister\cite7
\soulregister\ref7
	\renewcommand{\vo}[1]{{\textcolor{red}{\st{#1}}}}
    
\fi

\begin{document}

\title{Study of Silicon Photomultiplier Performance in External Electric Fields}
\newcommand{\IHEP}{a}
\newcommand{\SLAC}{b}
\newcommand{\Sherbrooke}{c}
\newcommand{\Carleton}{d}
\newcommand{\Indiana}{e}
\newcommand{\Erlangen}{f}
\newcommand{\PNL}{g}
\newcommand{\Duke}{h}
\newcommand{\Illinois}{i}
\newcommand{\ITEP}{j}
\newcommand{\SD}{k}
\newcommand{\LLNL}{l}
\newcommand{\RPI}{m}
\newcommand{\McGill}{n}
\newcommand{\Triumph}{o}
\newcommand{\IME}{p}
\newcommand{\Colorado}{q}
\newcommand{\BNL}{r}
\newcommand{\Laurentian}{s}
\newcommand{\Stanford}{t}
\newcommand{\NCarolina}{u}
\newcommand{\Drexel}{v}
\newcommand{\ORNL}{w}
\newcommand{\UMass}{x}
\newcommand{\Muenchen}{y}
\newcommand{\Bama}{z}
\newcommand{\Yale}{aa}
\newcommand{\Stony}{ab}
\newcommand{\IBS}{ac}
\newcommand{\UCalifornia}{ad}
\newcommand{\LHEP}{ae}

\author[\IHEP]{X.L.~Sun}
\author[\IHEP,1]{T.~Tolba}
\author[\IHEP,1,2]{G.F.~Cao}
\note[1]{Corresponding author}
\note[2]{also at University of Chinese Academy of Sciences, Beijing 100049, China}
\author[\IHEP]{P.~Lv}
\author[\IHEP]{L.J.~Wen}
\author[\SLAC]{A.~Odian}
\author[\Sherbrooke]{F.~Vachon}
\author[\Carleton]{A.~Alamre}
\author[\Indiana]{J.B.~Albert}
\author[\Erlangen]{G.~Anton}
% ...

\collaboration{nEXO collaboration}
\arxivnumber{ }
\date{\today}

\abstract{%
We report on the performance of silicon photomultiplier (SiPM) light sensors operating in electric field strength up to \SI{30}{\kilo\volt\per\cm} and at a temperature of \SI{149}{\kelvin}, relative to their performance in the absence of an external electric field. The SiPM devices used in this study show stable gain, photon detection efficiency, and rates of correlated pulses, when exposed to external fields, within the estimated uncertainties. No observable physical damage to the bulk or surface of the devices was caused by the exposure.
}


\maketitle


\section{Introduction}
\label{sec:sec-intro}
Silicon photomultipliers (SiPMs) are multi-pixel semiconductor devices, with pixels (microcells) arranged on a common silicon substrate~\cite{BONDARENKO2000}. Each microcell is a Geiger-mode avalanche photodiode (GM-APD), working above the breakdown voltage ($\it{U_\textrm{bd}}$), and a resistor for passive quenching of the breakdown. SiPMs are designed to have high gain (typically $\sim$~10$^{6}$), high photon detection efficiency ($\it{PDE}$)~\cite{nEXOSiPMpaper}, excellent time resolution, and wide range spectral response. They can be used to detect light signals at the single photon level and their dark noise rate can be significantly suppressed at low temperatures. Compared with traditional photomultiplier tubes (PMTs), SiPMs are more compact, can be produced with lower radioactivity~\cite{Ostrovskiy-SiPM2015}, and do not require high operating voltage.  These features make SiPMs very attractive photosensors for low-background, cryogenic applications such as the liquefied noble-element detectors used in dark matter searches and neutrino physics.

The next generation detector of the Enriched Xenon Observatory, nEXO~\cite{PhysRevC.97.065503,nEXOpCDR}, is one such application that intends to use SiPMs as photosensors to detect scintillation light from liquid xenon (peak wavelength at \SI{175}{\nano\meter} \cite{FUJII2015293}). nEXO will use isotopically enriched liquid xenon (LXe) in a time projection chamber (TPC) to search for the neutrinoless double beta decay of $^{136}$Xe. It has a projected half-life sensitivity of approximately $10^{28}$ years \cite{PhysRevC.97.065503}, covering most of the parameter space corresponding to the inverted neutrino mass hierarchy (assuming the standard mechanism for the decay and no axial coupling constant quenching). In order to achieve the required collection efficiency and energy resolution, a $\sim$~\SI{4}{\meter}$^{2}$ array of SiPMs will be placed on the cylindrical barrel of the TPC behind the field shaping rings. Such an arrangement will expose the SiPM devices to external electric fields of varying strength, depending on their position along the drift axis of the detector. Preliminary COMSOL~\cite{comsol} electrostatic simulations show that the SiPMs will be exposed to electric fields as high as \SI{20}{\kilo\volt\per\cm} in regions close to the cathode, with the electric field vector roughly perpendicular to the SiPM front surface. 

While it is known that SiPMs are insensitive to external magnetic fields \cite{SiPM_MagF}, their performance in strong external electric fields has not been extensively studied. The performance of a SiPM can be affected if the external electric field penetrates into the device. For example, if the external field reaches the avalanche region, typically located \SI{100}{\nano\meter} to \SI{500}{\nano\meter} below the semiconductor surface, it can change the breakdown initiation probability of a microcell~\cite{SPAD_breakdown}, thus affecting the photodetection efficiency. However, the top contact of each microcell is highly doped silicon, which should shield the inner volume of the microcell from the external fields. A suitable comparison is a Metal-Oxide-Semiconductor (MOS) capacitor~\cite{MOS-FET}, in which the electric field is screened either by a depletion or an accumulation region, depending on the semiconductor doping type and field direction. In the depletion regime, the field extends inside the semiconductor for a few nanometers for high doping concentrations (on the order of $10^{19}~\textrm{cm}^{-3}$) and hundreds of nanometers for low doping concentrations ($<10^{17}~\textrm{cm}^{-3}$). Because the internal structures and doping concentrations of SiPMs are not disclosed by vendors, the penetration depth of external electric fields is difficult to estimate a priori. Thus the field dependence of each type of device needs to be measured experimentally.

Reference~\cite{Anderson:2017zun} demonstrates that at room temperature SiPMs can operate in a \SI{3.2}{\kilo\volt\per\cm} external electric field with no discernible performance change. In this paper, we study the behavior of SiPMs in a cryogenic environment and with electric fields up to \SI{30}{\kilo\volt\per\cm}. This study measures the changes in the SiPM performance parameters, such as gain, correlated noise, and photon detection efficiency.

\section{Instrumentation}
\label{sec:Inst}

\subsection{The Test Station}
\label{sec:test_station} 
\begin{figure}
    \centering
   \includegraphics[width=0.9\linewidth]{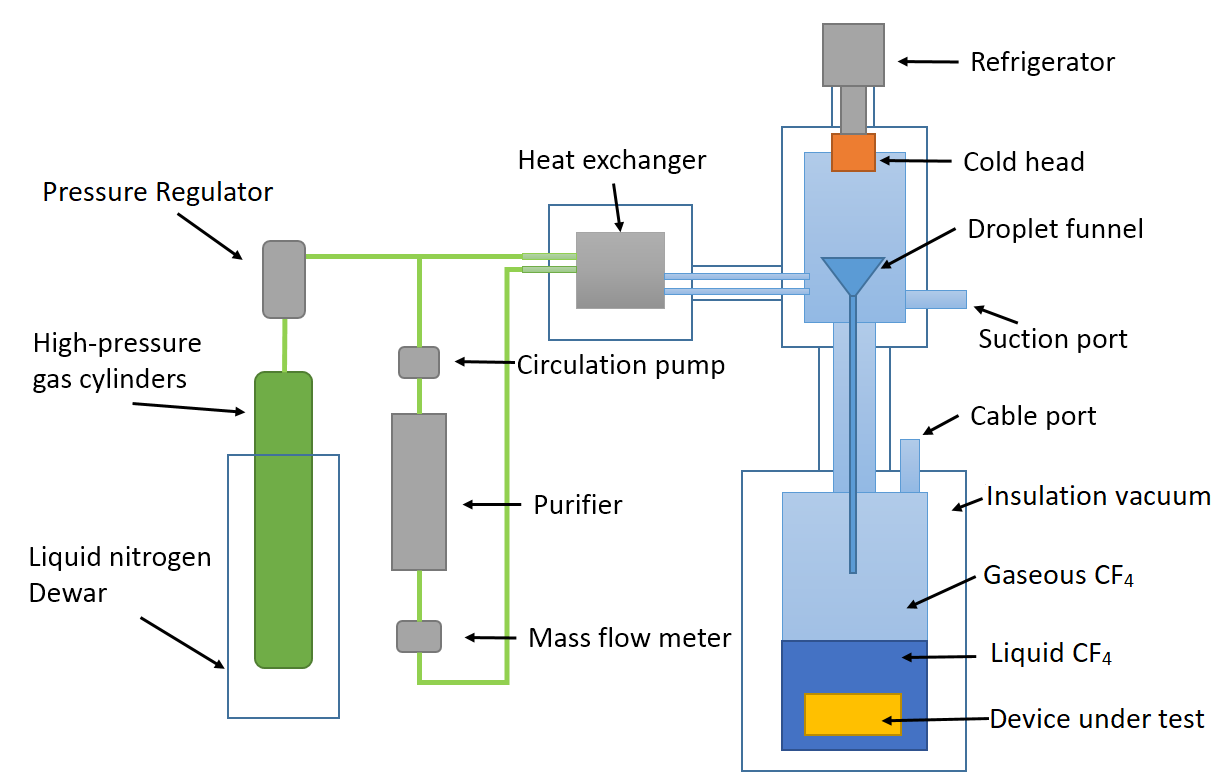}
    \caption{Schematic diagram of the cryogenic test station operated with liquefied CF$_{4}$ in these measurements.} 
    \label{fig:Test_station}
\end{figure}

Figure~$\ref{fig:Test_station}$ shows a schematic diagram of the cryogenic test station located at the Institute of High Energy Physics (IHEP) in Beijing, China. The inner test chamber is \SI{200}{\milli\meter} in diameter and \SI{340}{\milli\meter} in length. The inner chamber is vacuum-insulated ($\sim$~10$^{-5}$ \SI{}{\milli\bar}) and wrapped with thermal insulation sheets. Although the station is built to run with LXe, it is capable of handling other gases such as Ar and CF$_{4}$. The liquid level inside the test chamber is monitored by eight platinum resistance temperature detectors (RTD) placed at specific heights along a calibrated nylon rod.

\begin{figure}
    \centering
   \includegraphics[width=0.8\linewidth]{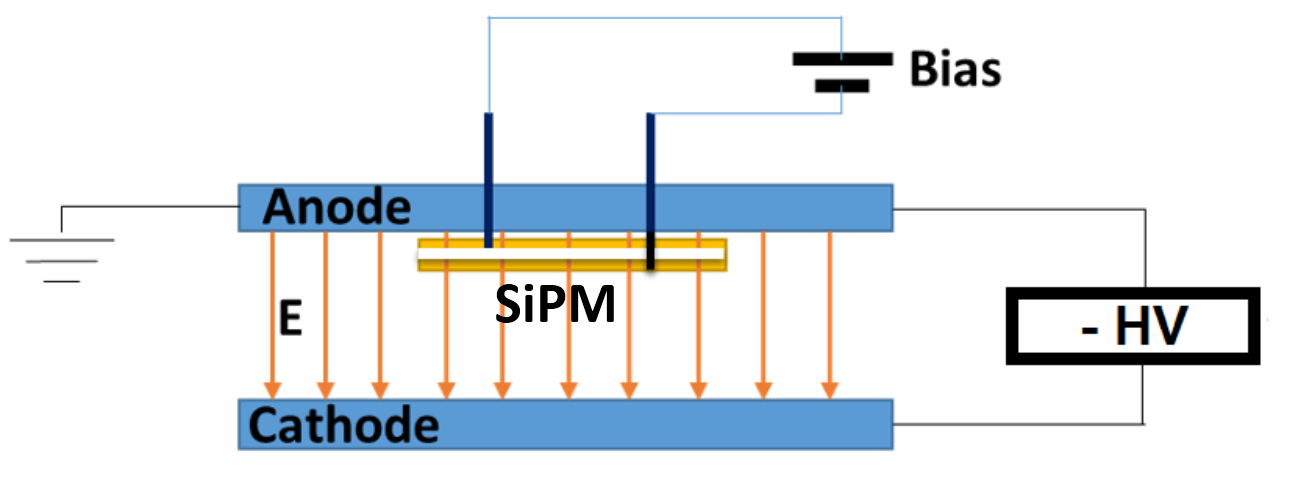}
    \caption{Conceptual design of the SiPM high electric field performance test.} 
    \label{fig:Test_design}
\end{figure}

\begin{figure}
    \centering
    \includegraphics[width=0.5\linewidth]{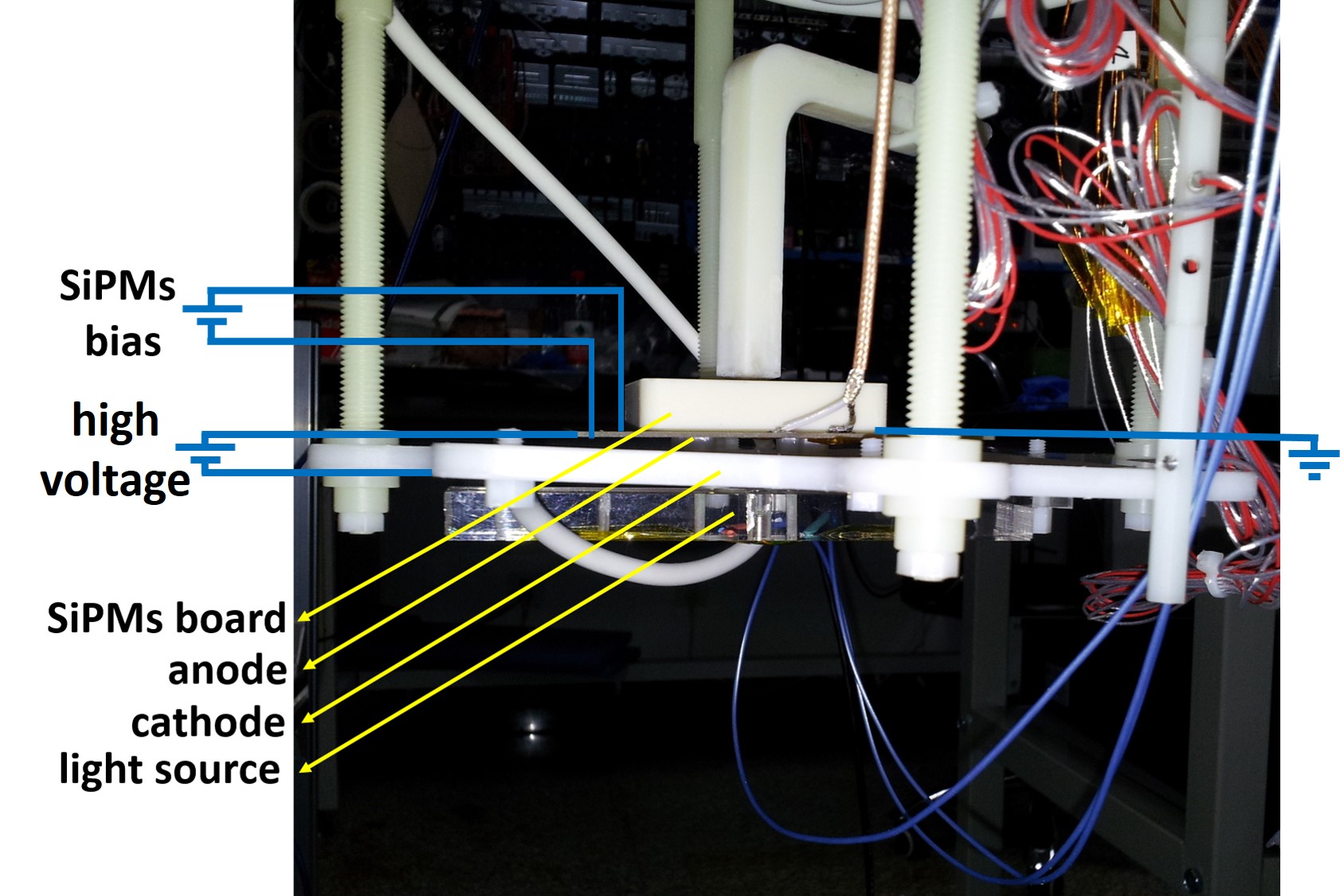} \includegraphics[width=0.415\linewidth]{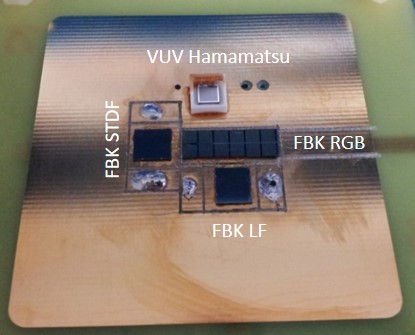} 
    \caption{(Left) The assembly of the test setup: SiPMs, cathode and light source in the inner chamber. (Right) The SiPMs used in this study attached to the grounded metallic sheet.}
    \label{fig:SiPMs_array}
\end{figure}

A schematic of the SiPM test setup is shown in Figure~$\ref{fig:Test_design}$, with pictures of the assembly and the SiPM devices shown in Figure~$\ref{fig:SiPMs_array}$. The SiPMs are mounted on a support board and are placed between a high voltage (HV) cathode mesh and a grounded anode. 

For accurate vertical positioning of the SiPMs with respect to the cathode, and to ensure that the SiPM-cathode distance is stable during the measurement, \SI{5}{\milli\meter} long nylon spacers were placed between the SiPM support board and the cathode mesh. The thickness of the SiPMs was measured to be \SI{0.5}{\milli\meter}, such that the distance between cathode and SiPM surface was \SI{4.5}{\milli\meter}. The center of the SiPMs support board was aligned to the center of the cathode mesh. The positioning of the SiPMs near the center of the electrode plates and their relatively small size (see Table~\ref{tab:SiPM_prope}) compared to the cathode mesh (\SI{100}{\milli\meter} x \SI{100}{\milli\meter}), ensures the uniformity of the external electric field across the surface of the SiPMs. The SiPM bias and signal connections pass through the anode board. The cathode mesh is placed in a nylon frame to avoid discharges between the edge of the cathode plate and the side walls of the inner chamber. All internal components close to the cathode are made of nylon as insulating material. A Spellman high voltage power supply~\cite{HVmodule} was used to bias the cathode to a maximum of \SI{-20}{\kilo\volt}, with \SI{100}{\volt} precision. With this setup we have focused on tests where the external fields are perpendicular to, and pointing away from, the device front surface, essentially the nEXO configuration.

The entire assembly, including the anode and the cathode, is submerged in liquefied CF$_{4}$ (LCF$_{4}$) at $\sim$~\SI{149}{\kelvin} and a pressure of $\sim$~1.4 atm. 
This measurement was carried out in LCF$_{4}$ as a practical alternative to LXe. The relevant electrical and thermal properties of LCF$_{4}$ are close to those of LXe, as the LCF$_{4}$ static dielectric constant (1.614 at \SI{149}{\kelvin}~\cite{CF4}) is only $\sim$~\SI{15}{\percent} lower than that of LXe (1.874 at \SI{165}{\kelvin}~\cite{Amey:1964}) and its boiling point (\SI{145}{\kelvin} at 1 atm)  is only $\sim$~\SI{12}{\percent} lower than that of the LXe (\SI{165}{\kelvin} at 1 atm) \cite{CF4-mp}. Because of the use of CF$_{4}$, this study does not explore the characteristics of LXe breakdown relevant to nEXO.

\subsection{The Photo-Sensors}
\label{sec:SiPMs}
In this study three SiPM devices of interest to the nEXO collaboration were tested. Two devices were from FBK (Fondazione Bruno Kessler)~\cite{fbk}; FBK RGB-HD (FBK-RGB) and FBK Vacuum Ultra Violet (VUV) low field (FBK-LF), and one device was from Hamamatsu~\cite{hamamatsu}; Hamamatsu VUV3 (detailed information for each device is given in Table~$\ref{tab:SiPM_prope}$). In addition to the two FBK devices listed in Table~$\ref{tab:SiPM_prope}$, an FBK standard field (FBK-STDF) SiPM was also considered. However, the operating voltage of the FBK-STDF device is significantly reduced at cryogenic temperatures, leading to poor performance, and the results are therefore not reported in this publication. 

The FBK devices came as bare dies while the SiPM from Hamamatsu was packaged in a ceramic frame. The FBK-RGB device has the cathode and anode electrodes located on the front and back surfaces, respectively, while they are in the opposite configuration for the FBK-LF SiPM. Thus the two FBK devices are subject to opposite relative orientations of the internal electric field with respect to the external electric field. A thin layer of copper substrate with several isolated pads, attached to a PCB circuit board, was used to mount the devices. The FBK devices were attached to the metal substrate by conductive silver glue, and the cathodes were connected to contacts on isolated pads on the metal substrate by wire bonding. The wire bonds were then protected by UV curable adhesive. The two pins of the Hamamatsu device passed through holes on the substrate. All connections to preamplifier were made from the back of the substrate.

\begin{table}
  \begin{center}
    \caption{Characteristics of the SiPM devices used in this study.}
    \label{tab:SiPM_prope}
    \begin{tabular}{c|c|c|c|c|c|c}
      \textbf{Device} & \textbf{Dimentions} & \textbf{Fill} & \textbf{Breakdown} & \textbf{Over} & \textbf{VUV} & \textbf{Packing}\\
       & & \textbf{factor} & \textbf{voltage} & \textbf{voltage} & \textbf{sensitive} & \\
       &  \textbf{[mm$^{2}$]} & \textbf{[\% ]} & \textbf{@149 K} & \textbf{($\it{U_\textrm{ov}}$)} & & \\
      \hline
      FBK VUV  Low Field & 5.96 x 5.56 & 73.0 & 29.0 V & 4.5 V & Yes & Die \\
      FBK RGB-HD \cite{Serra2013} & 15.30 x 4.95 & 72.5 & 24.0 V & 3.0 V & No & Die\\
      Hamamatsu VUV3 & 3.40 x 3.40 & 50.0 & 44.0 V & 3.5 V & Yes & Ceramic mount\\
    \end{tabular}
  \end{center}
\end{table}
     
\subsection{The Data Acquisition (DAQ) System and Data Collection}
\label{sec:DAQ}
\begin{figure}
\centering
\includegraphics[width=0.98\linewidth]{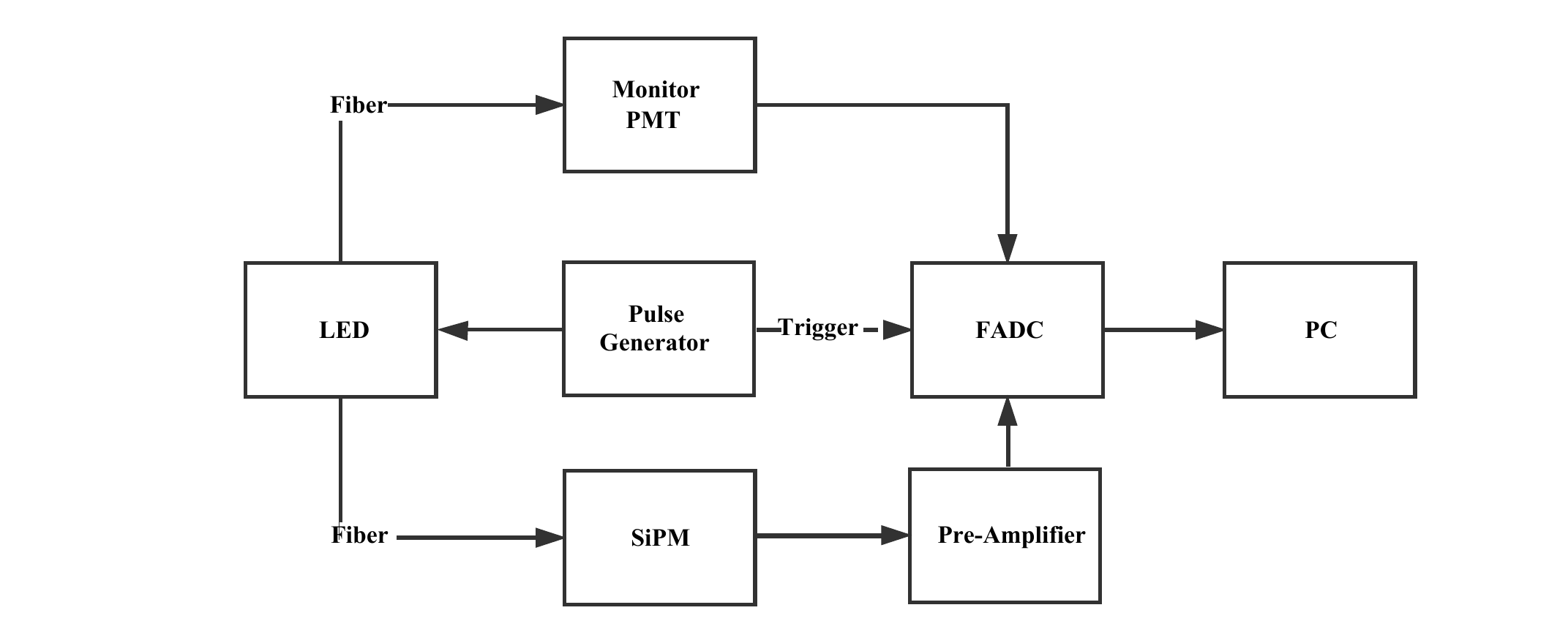}
\caption{Schematics of the DAQ System for the SiPM measurements with an LED.}
\label{fig:DAQ}
\end{figure}

Figure~$\ref{fig:DAQ}$ shows a schematic diagram of the DAQ system used for the measurement under illumination with LED light. Two different models of preamplifier (which also deliver the bias voltage to the SiPMs) were used in this experiment. For the Hamamatsu VUV3 SiPM we used a commercial Photonique AMP 0604 preamplifier \cite{preamp}, placed outside the chamber at room temperature. For the FBK devices, two custom-made preamplifier were placed inside the chamber in the cold CF$_{4}$ gas volume. The two types of preamplifier have very similar properties: current sensitive with $4-10$ V supply voltage, signal rise time $\sim$~\SI{5}{\nano\second}, relatively large bandwidth $10-40$ MHz, and gain between 10 and 40. The preamp signals from the SiPMs were digitized by a $4-$channel CAEN DT5751 digitizer unit~\cite{caen} with 10 bit resolution, \SI{1}{\volt} dynamic range, and a maximum sampling rate of \SI{1}{\giga\hertz}. The captured pulses were sent to a PC and recorded by custom LabView software~\cite{labview} for further off-line processing with ROOT \cite{ROOT}. 

Two sets of data with different illumination conditions were collected at external electric field strengths of 0, 6, 12, 18, 24 and \SI{30}{\kilo\volt\per\cm}. The first data set was collected in the dark with no LED light illumination. In this case, data collected from two different runs were analyzed: one for the Hamamatsu device and one for the FBK devices. The signals from the SiPMs were acquired by setting a threshold on the SiPM output signal amplitude using a leading edge discriminator. The acquisition time window was set to \SI{200}{\micro\second} for the Hamamatsu SiPM and \SI{1}{\micro\second} for the FBK devices. The data was taken over a period of 30 minutes for the Hamamatsu SiPM, with an average trigger rate of \SI{10}{\hertz}. For the FBK SiPMs, the maximum number of events was set to 5x10$^{5}$ and 2x10$^{4}$ for the FBK-RGB and FBK-LF devices, respectively. For the FBK-LF device, the data collected at \SI{6}{\kilo\volt\per\cm} was not saved, hence this data point is not present in the following results.

\begin{figure}
   \centering
   \includegraphics[width=0.9\linewidth]{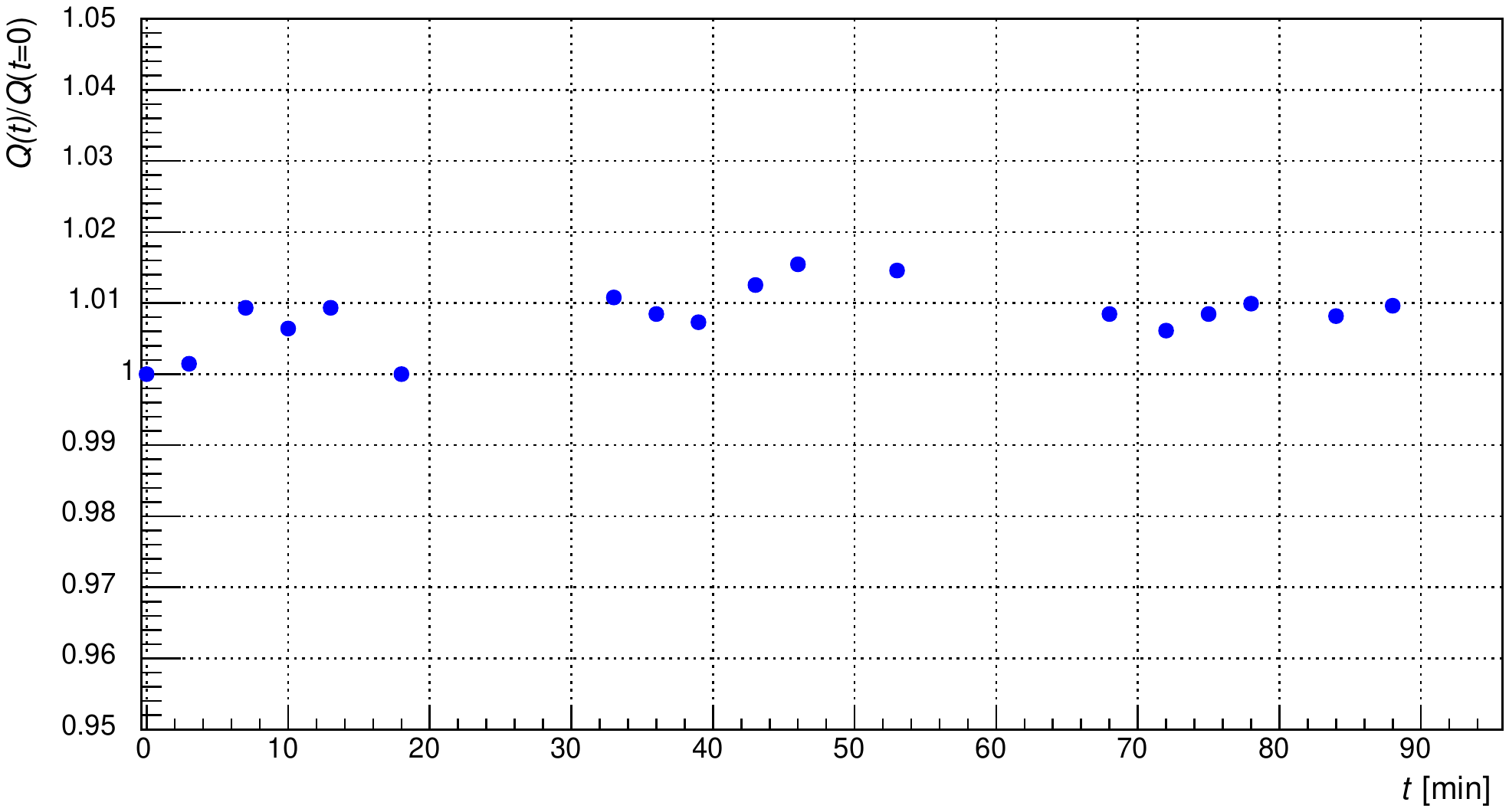}
   \caption{The total charge collected by the monitor PMT as a function of measurement time when illuminated by the LED light source.}
   \label{fig:PMT_stability}
\end{figure}

The second data set was collected using a blue (\SI{465}{\nano\meter}) LED that was placed outside the test (LCF${_4}$) chamber at room temperature and driven by a pulse generator at \SI{10}{\kilo\hertz} frequency. The light pulse was split into two channels and delivered by optical fibers to the inner test chamber behind the cathode mesh (facing the SiPM) and a PMT outside the chamber used to monitor the stability of the LED light. The PMT used in this test was an ET Enterprises 9364UFLB PMT~\cite{PMT} with a specified gain of 1.3x10$^{7}$. In this data set both the SiPM and the PMT signals were triggered by the pulse generator. The output signals of the monitor PMT were sent directly to the digitizer; no preamplifier was used. The data acquisition program collected the waveforms with a \SI{2}{\micro\second} readout window and the maximum number of events was set to 5x10$^{4}$. 
The instability in the output charge of the monitor system (PMT-LED combination) was studied as a function of the measurement time and found to be $\sim$~\SI{2}{\percent} with respect to that at the beginning of the measurement, as shown in Figure~$\ref{fig:PMT_stability}$ (the statistical error bars, much less than \SI{1}{\percent}, are not visible on this scale). 

For both data sets, all experimental configurations were kept the same throughout the data taking at various external electric fields.

\section{Results}
\label{sec:results}
\subsection{Pulse Finding Algorithm and Systematic Uncertainty}
\label{sec:PFA}
\begin{figure}
    \centering
    \includegraphics[width=0.9\linewidth]{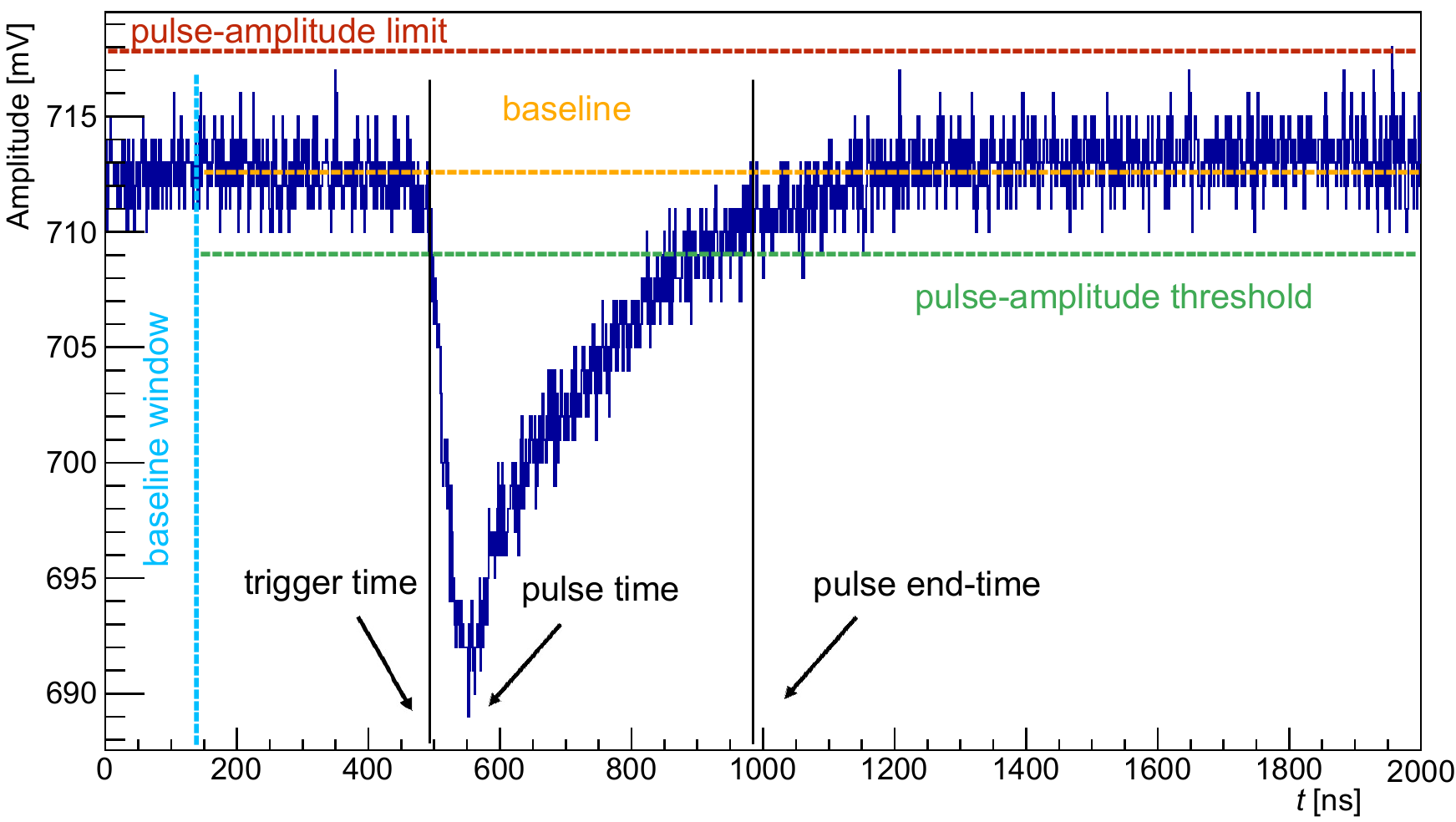}
    \caption{An example of a typical digitized waveform shown together with its baseline, trigger threshold, trigger time and pulse end times, as defined in the text. }
    \label{fig:Pulse_Selection}
\end{figure}

\begin{figure}
    \centering    
 \includegraphics[width=0.6\linewidth]{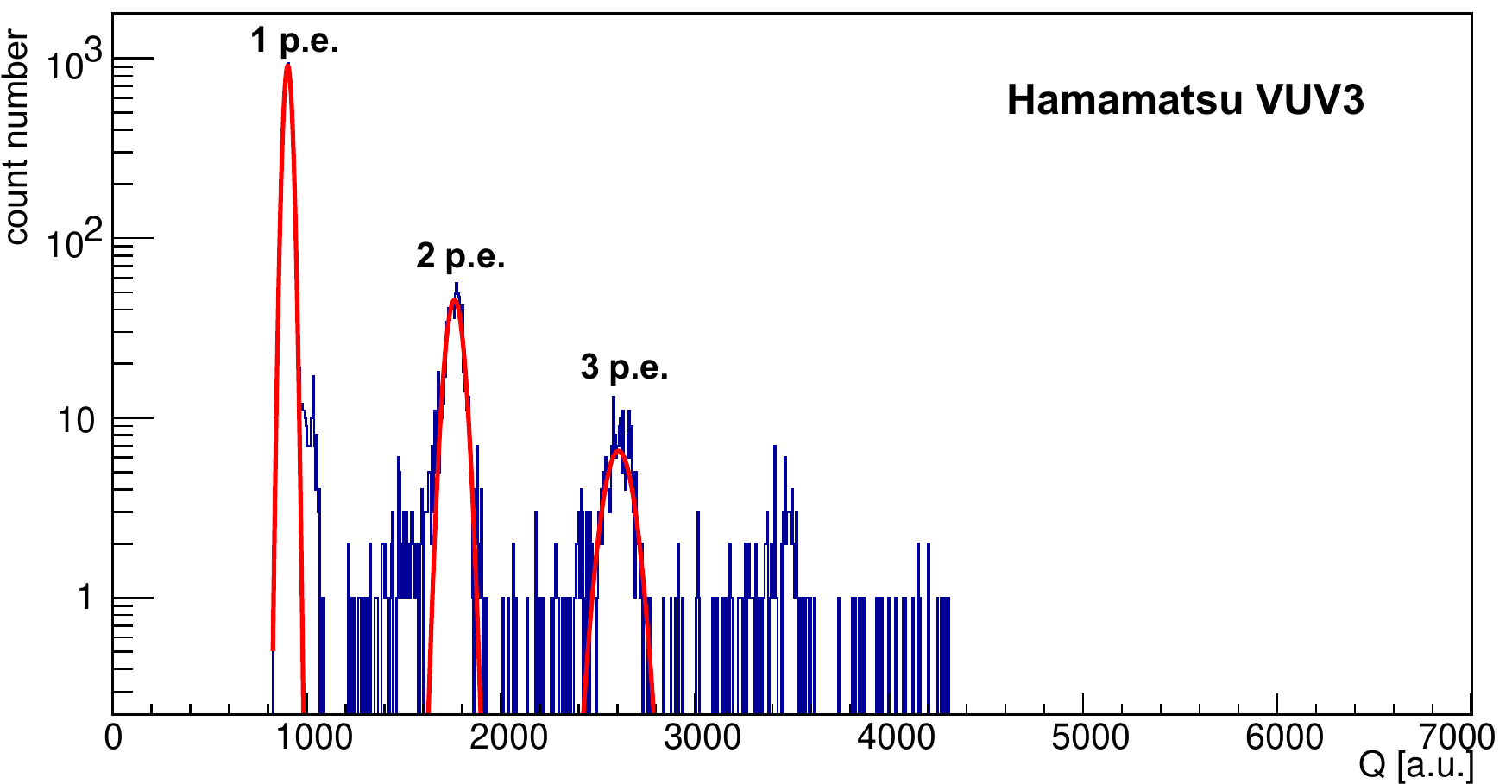}
 \includegraphics[width=0.6\linewidth]{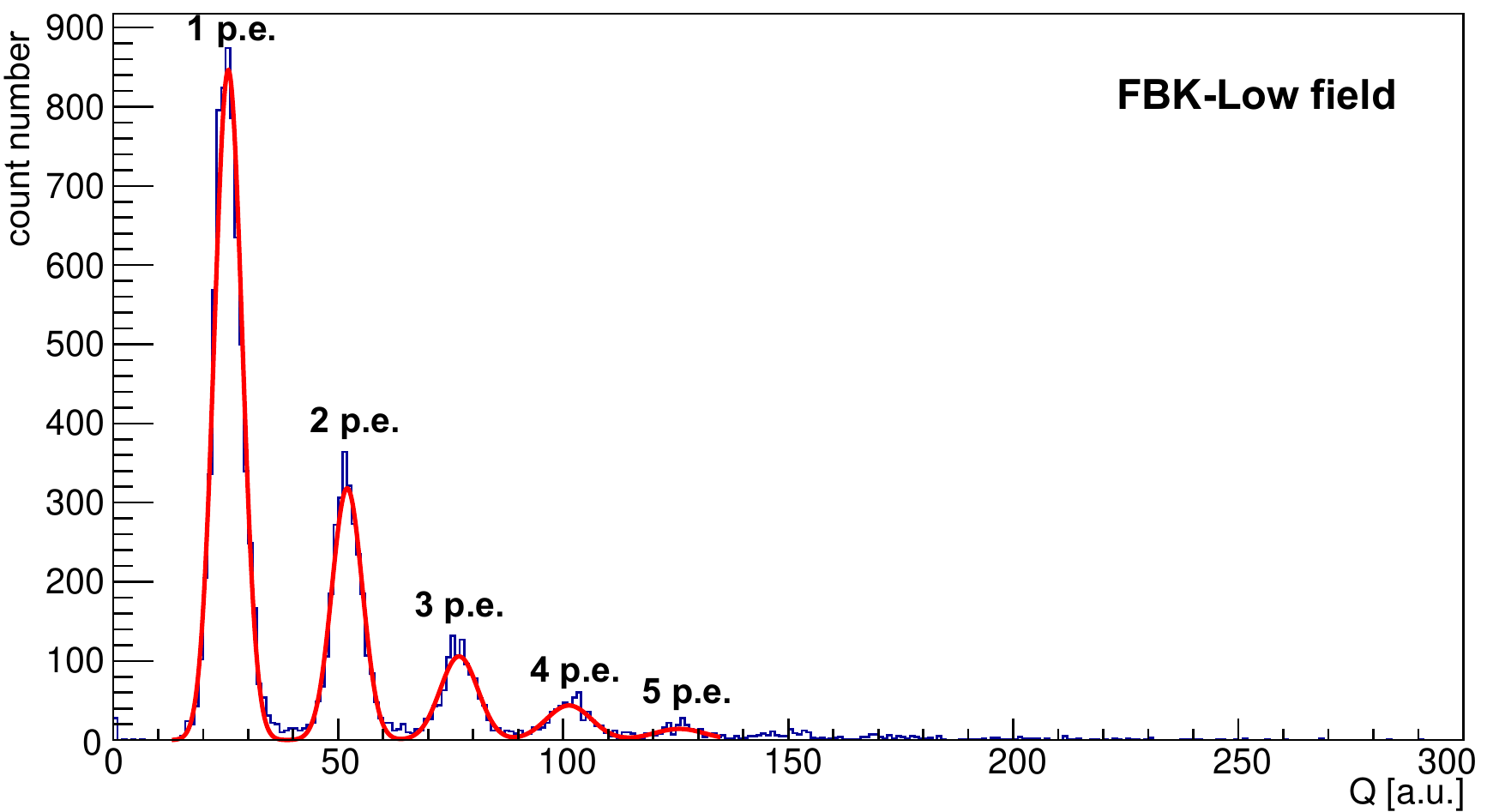}
 \includegraphics[width=0.6\linewidth]{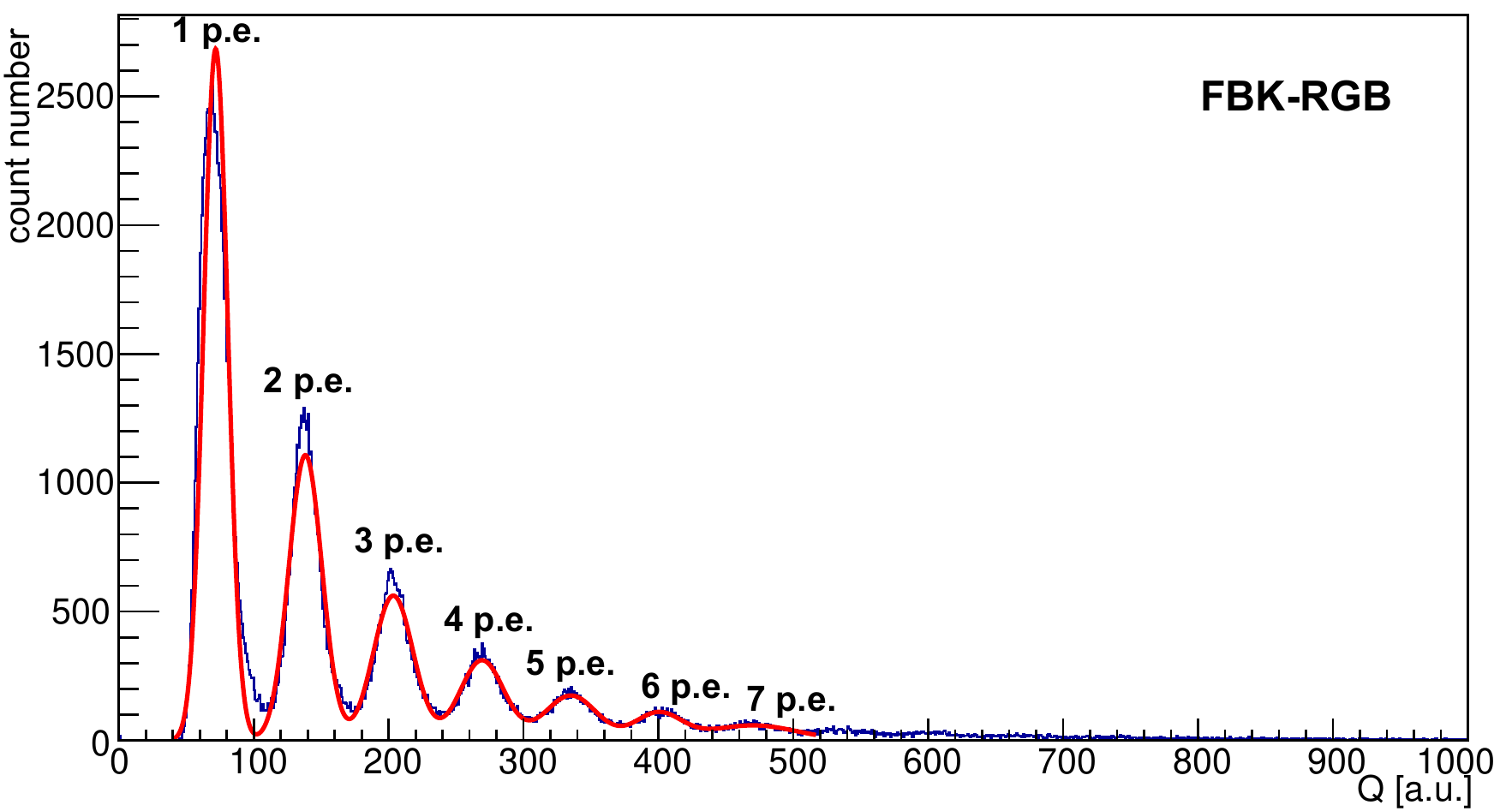}
 \caption{The output charge spectrum of the Hamamatsu VUV3 (top), FBK-LF (middle) and FBK-RGB (bottom) SiPMs in an external electric field of $\it{E}$ = \SI{18}{\kilo\volt\per\cm}. The multi p.e.\ peaks are fitted with a sum of independent Gaussian functions (red line). The shoulder on the right side of the single p.e. peak is due to after-pulses.}
\label{fig:Single_photon}
\end{figure}

The offline analysis first discriminates light signals from noise using a pulse finding algorithm (PFA), then calculates the total charge, $\it{Q}_{tot}$, collected by the light sensors (the SiPMs or the PMT) for each event. The PFA algorithm selects signal pulses using a series of cuts. It first sets a lower limit on the pulse amplitude above the baseline. It then looks at the correlation between the pulse width and the corresponding integrated charge, as a 2D histogram. A ROOT graphical cut is used in this 2D plot to select and exclude the false signals and noise pulses with low charge and/or small width. The signal selection and noise rejection efficiencies of the cuts are measured from the individual spectrum of each cut.  Close to \SI{100}{\percent} of the signals pass the cuts while more than $\sim$~\SI{95}{\percent} of noise events are rejected. The total charge for each identified signal is calculated by integrating the total ADC values in the pulse after baseline subtraction. The baseline is defined as the average of the waveform in the time window prior to the trigger ([0, $140-500$] ns, depending on the position of the light signal in the waveform for each SiPM). The analysis is repeated for the data collected at different external electric fields. Figure~$\ref{fig:Pulse_Selection}$ shows an example of a typical waveform along with some of its characteristic features, while Figure~$\ref{fig:Single_photon}$ shows the output charge spectrum for the three SiPMs at $E$ = \SI{18}{\kilo\volt\per\cm}. Each peak above zero in Figure~$\ref{fig:Single_photon}$ corresponds to a quantized number of photoelectrons, p.e., while the few entries around zero are caused by the inefficiency of the PFA noise rejection. These noise events are not included in the following calculations. The multi p.e.\ peaks are fitted with a sum of independent Gaussian functions to estimate the relative gain.

Since this study mostly involves relative measurements, systematic effects that are independent of the external electric field cancel out. For the gain measurements, the total uncertainty is dominated by the systematic uncertainty related to changes in electronics pickup from the HV power supply as a function of high voltage. To estimate this uncertainty we measured the FWHM of the baseline variations at different high voltage settings. We found at most $\sim$~\SI{7}{\percent} deviation, for widths measured at non-zero external electric fields compared to that at zero external field. This baseline noise was then added to simulated signal pulses to estimate its effect on the total charge calculation. This Monte Carlo study indicates that the excess noise pickup can lead to systematic errors in the measurement of the gain by the PFA by up to \SI{1.5}{\percent}. The statistical uncertainty for the gain measurement was found to be negligible.   

The correlated noise measurements, on the other hand, are limited by statistical uncertainties, as the following two systematic uncertainties are found to be sub-dominant. The first is interference generated by the high voltage supply. This noise usually appears as symmetrical pulses around the baseline and can be excluded efficiently by setting a limit on the positive amplitude of the waveform in the PFA algorithm. A second source of systematic uncertainty is the presence of background light signals due to possible faint discharges in the liquid at high electric fields.  We measured this by looking for additional signals in the first $\mu$s after the trigger, but no such excess of light signals was observed. The total systematic uncertainty is estimated to be below \SI{1}{\percent} for the correlated noise measurements. Table~\ref{tab:Error_table} summarizes the most significant uncertainty sources for each of the studied parameters. 

To account for the uncertainty on the vertical distance between the anode, SiPMs, and cathode surfaces, a \SI{5}{\percent} uncertainty in the value of the electric field strength was assumed.

\begin{table}
  \begin{center}
    \caption{Summary of the most significant error sources to the different parameters studied in this paper.}
    \label{tab:Error_table}
    \begin{tabular}{l|l|c|l}
      \textbf{Parameter} & \textbf{Dominant error} & \textbf{Value [\%]} & \textbf{Source of the uncertainty}\\
      \hline
      Relative gain & Systematic & $\sim$~1.5 & HV induced baseline noise \\
      Prompt corss-talk probability & Statistical & $\sim$~2 & Total number of prompt signals\\
      Delayed correlated noise probability & Statistical & $\sim$~1.5 - 4 & Counts of the delayed CN signals\\
      Relative \textit{PDE} & Systematic & $\sim$~2 & PMT-LED output stability\\
    \end{tabular}
  \end{center}
\end{table}

\subsection{Results for Data Collected in the Dark}
\label{sec:Gain_dark}
We used the data set collected in the dark to study the relative gain and the stability of the correlated noise at different external electric fields. Because of the large DAQ dead-time, we were not able to make a precise measurement of the dark rate in this study. 

\subsubsection{Relative Gain}
\label{sec:RGain}
The gain of a SiPM can be defined as the mean number of output electrons in the single p.e.\ peak \cite{AdvanSiD}. We used the charge distribution of the prompt signal, e.g. in Figure~$\ref{fig:Single_photon}$, to study the relative stability in the gain of the SiPMs at different external electric fields. The mean value of each individual fitted Gaussian is used to estimate the average charge of the corresponding number of photoelectrons, $\it{Q_{n~p.e.}(E)}$. The slope of the $\it{Q_{n~p.e.}(E)}$ values, when plotted against the number of photoelectrons $\it{n}$, is then used to calculate the average charge of the SiPM single p.e.\ response at a specific $\it{E}$ value, $\bar{Q}(E)$. Thus the stability of the SiPM gain at different electric fields can be assessed by the ratio, $\eta_{Gain}$, of the charge amplitude, $\it\bar{{Q}}$($\it{E}$) (with $\it{E}$ = 6, 12, 18, 24, \SI{30}{\kilo\volt\per\cm}), to that in the absence of the external field:

\begin{equation}
\eta_{Gain} = \frac{\bar{Q}(E)}{\bar{Q}(E=0)}
\end{equation}

\begin{figure}
   \centering
   \includegraphics[width=0.9\linewidth]{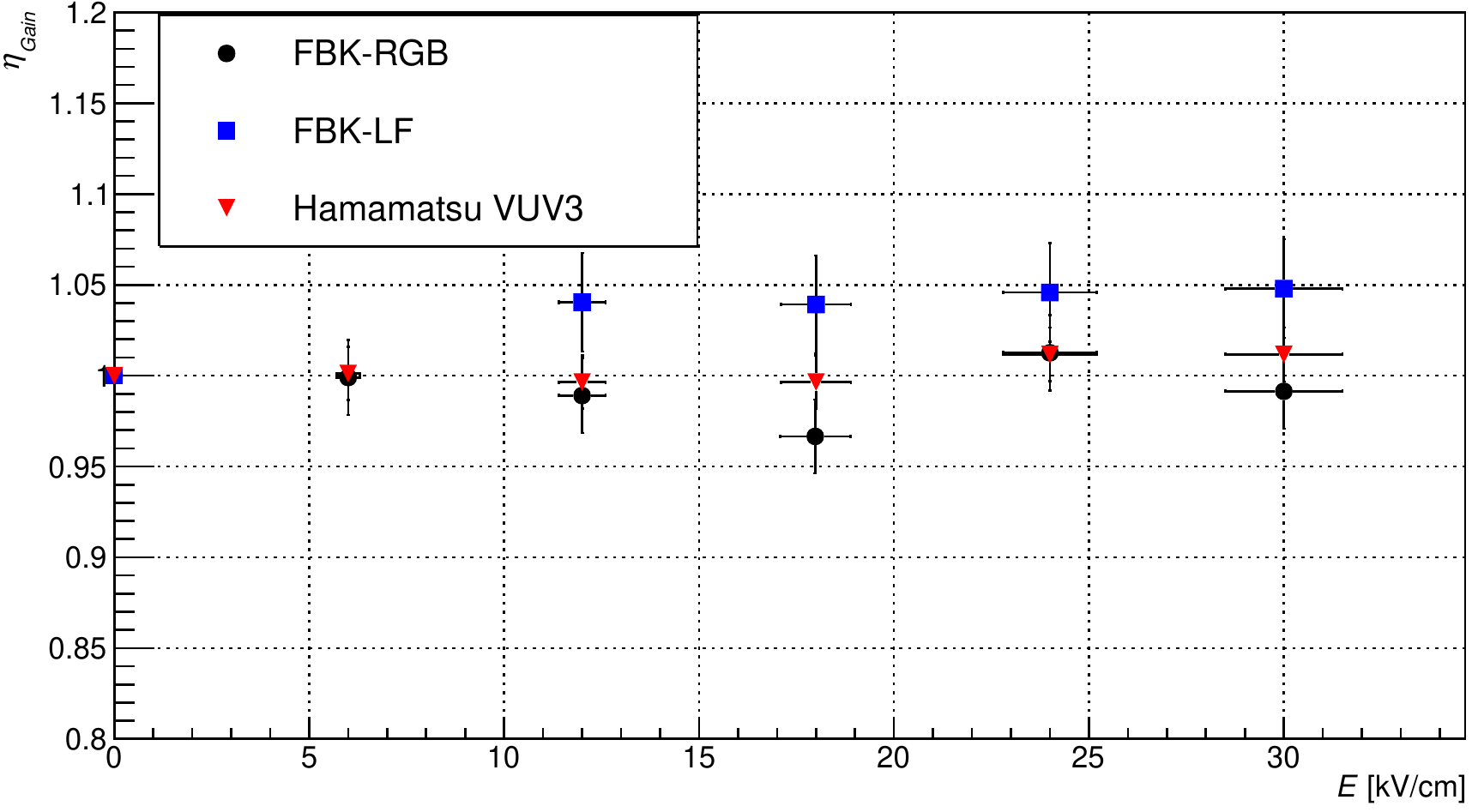}
   \caption{The SiPM relative gain measurements in the dark as a function of external electric field strengths for FBK-RGB (black circles), FBK-LF (blue squares) and Hamamatsu VUV3 (red triangles).}
   \label{fig:relative_gain_dark}
\end{figure}

Figure~$\ref{fig:relative_gain_dark}$ shows that the relative gain of all SiPMs stays constant as a function of the external electric field, with deviations less than $\sim$~\SI{5}{\percent} of the value in the absence of an external electric field, and all variations are consistent with the magnitude of the uncertainties.

\subsubsection{Prompt Cross-Talk Probability}
\label{sec:PCT}
\begin{figure}
   \centering
   \includegraphics[width=0.9\linewidth]{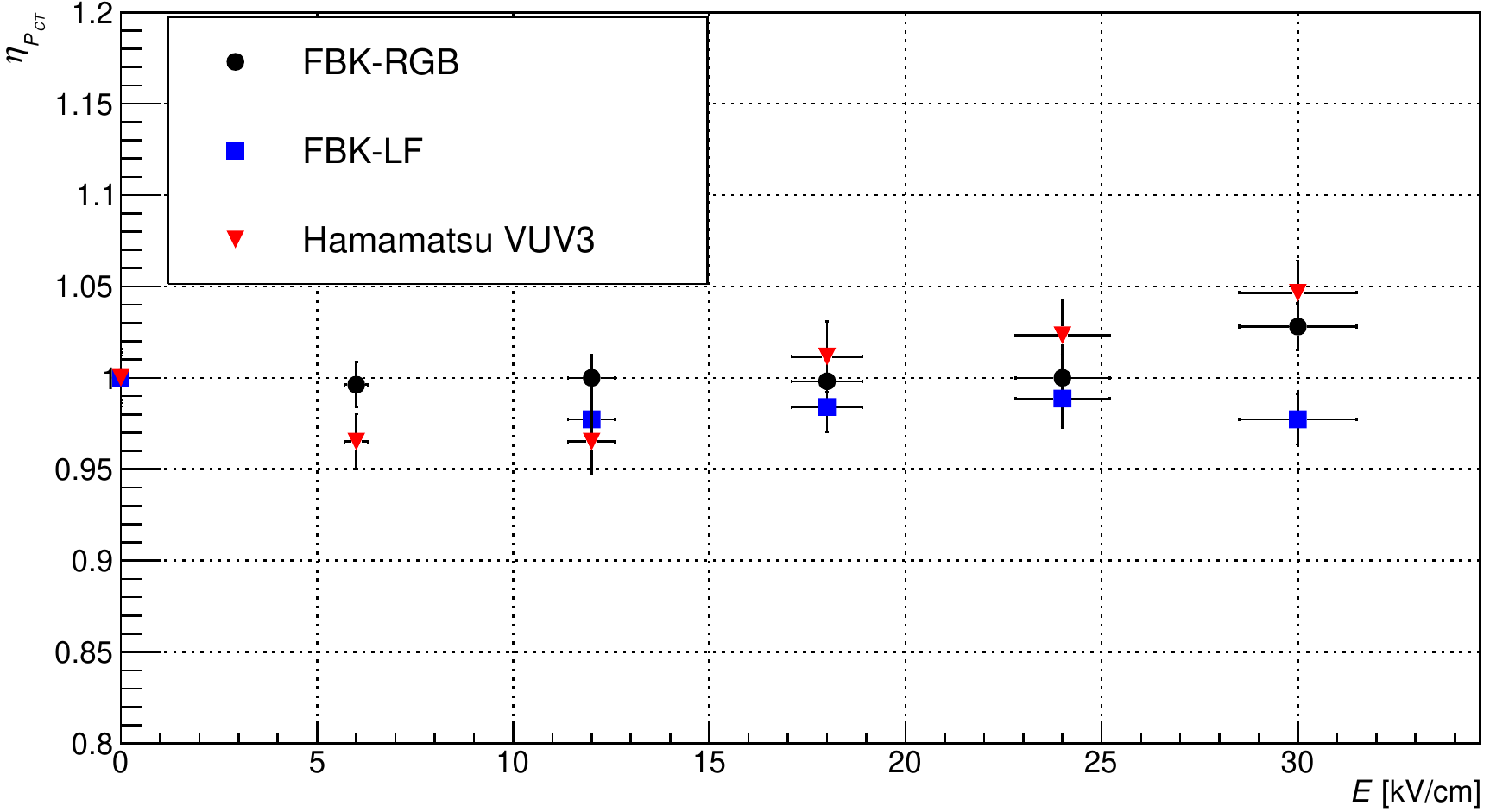}
   \caption{The relative prompt cross-talk, as a function of external electric field values for FBK-RGB (black circles), FBK-LF (blue squares) and Hamamatsu VUV3 (red triangles).}
   \label{fig:P_CT}
\end{figure}

Correlated signals are an important source of noise in SiPMs. They are composed of prompt optical crosstalk and delayed after-pulses~\cite{nEXOSiPMpaper}. The delayed correlated noise probability is discussed in section \ref{sec:PCN}. The origin of prompt crosstalk can be understood as follows: when undergoing an avalanche, carriers near the p-n junction emit photons, due to the scattering of the accelerated electrons. These photons tend to be at near infrared wavelengths and can travel substantial distances through the device, including to neighboring microcells where they may initiate secondary Geiger avalanches.  As a consequence, a single primary photon may generate signals equivalent to 2 or more photoelectrons~\cite{SensL}. The prompt crosstalk probability, $\it{P}_{CT}$, depends on over-voltage, $\it{U_\textrm{ov}}$, which is the excess bias beyond the breakdown voltage, device-dependent barriers for photons (trenches), and the size of the microcells. Assuming that in the dark accidental coincidences of multiple pulses triggered by dark noise is negligible at cryogenic temperatures, hence only 1-photoelectron equivalent signals are expected, the probability of prompt crosstalk can be calculated as:

\begin{equation}
P_{CT}=\frac{N_{> 1~ p.e.}}{N_{total}}
\end{equation}\
where $\it{N}$$_{> 1~ p.e.}$ is the number of the prompt signals with a measured charge of at least 1.5 p.e., and $\it{N}$$_{total}$ is the total number of prompt signals above noise. Figure~$\ref{fig:P_CT}$ shows $\eta_{P_{CT}}$, the ratio of $\it{P}_{CT}$($\it{E}$) to $\it{P}_{CT}$($\it{E=0}$), as a function of the external field. For the FBK devices $\eta_{P_{CT}}$ does not show a dependence on the external electric fields, within the uncertainty of our measurements. For the Hamamatsu VUV3 there may be a small dependence of $\eta_{P_{CT}}$ on the external field, but the effect is not significant considering the magnitude of the uncertainties. In all cases possible dependencies on the external field are below \SI{5}{\percent}. 

\subsubsection{Delayed Correlated Noise Probability}
\label{sec:PCN}
Both after-pulsing and delayed crosstalk events originate from an existing pulse. After-pulsing is due to the carriers trapped in silicon defects during the avalanche multiplication, then released later during the recharge phase of the microcell. Delayed crosstalk is generated by a similar mechanism to prompt crosstalk. The difference is that the photons generated during the avalanche process are absorbed in the inactive regions of the neighboring cells instead. It takes some time for the minority charge carriers to diffuse into the active region, causing a delayed signal~\cite{AdvanSiD}.  In our measurement, we cannot separate after-pulsing from delayed crosstalk and we count them together as delayed correlated noise.  

To estimate the delayed correlated noise probability, $\it{P}_{CN}$, we count the number, $N_{1 \mu s}$, of clearly separated pulses occurring immediately after the primary pulse but within a \SI{0.8}{\micro\second} and a \SI{1}{\micro\second} time interval after the trigger for the FBK devices and the Hamamatsu SiPM, respectively. The primary pulse time window is found to be $\sim$~\SI{30}{\nano\second} for the Hamamatsu device and $\sim$~\SI{20}{\nano\second} to \SI{55}{\nano\second} for the FBK devices. $\it{P}_{CN}$ is then estimated by normalizing $N_{1 \mu s}$ to the total number of events that contain prompt signals, $N_{prompt}$:

\begin{equation}
P_{CN}=\frac{N_{1\mu s}}{N_{prompt}}
\end{equation}

\begin{figure}
   \centering
   \includegraphics[width=0.9\linewidth]{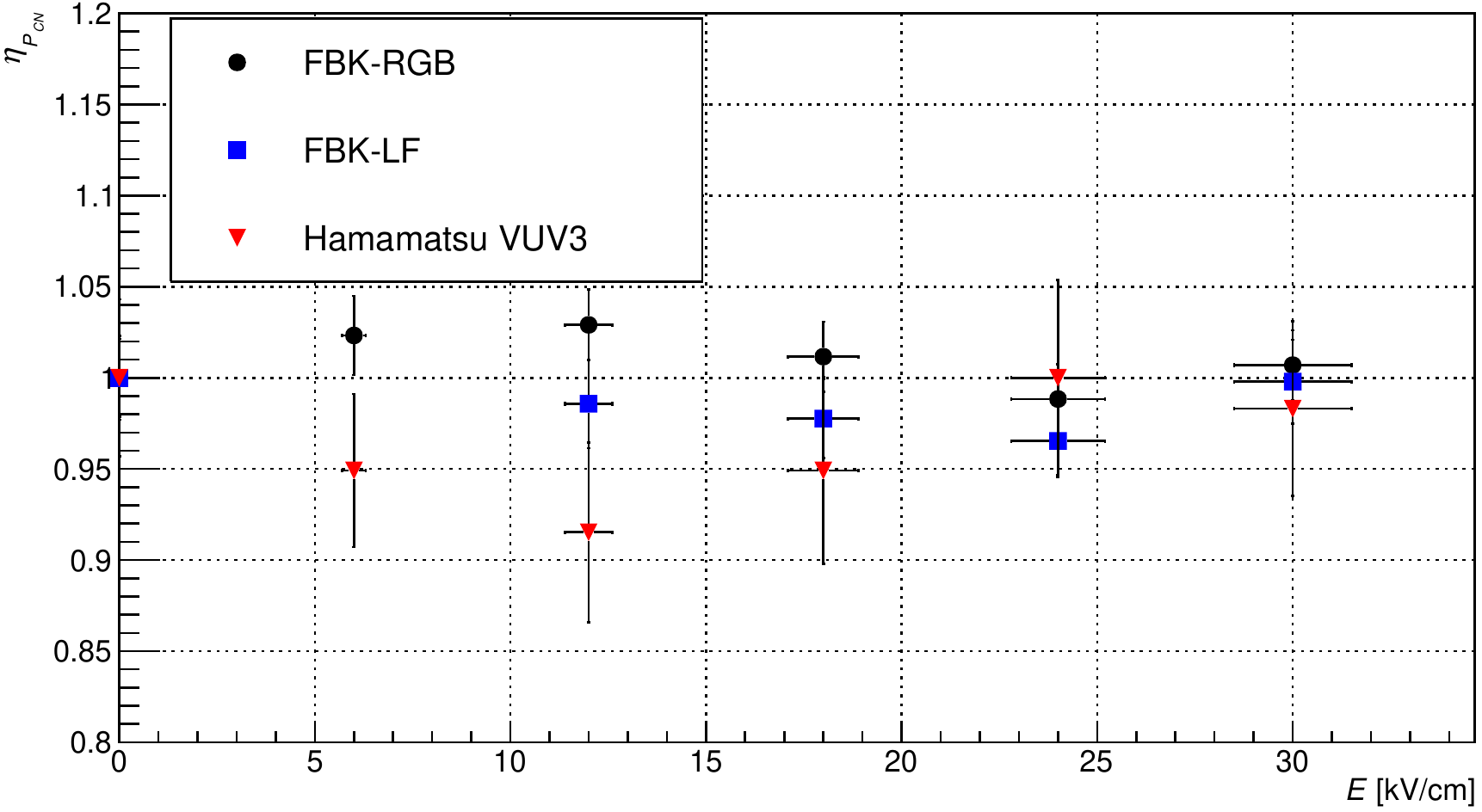}
   \caption{The relative delayed correlated noise as a function of external field values for FBK-RGB (black circles), FBK-LF (blue squares) and Hamamatsu VUV3 (red triangles).}
   \label{fig:P_CN}
\end{figure}

At the $\it{U_\textrm{ov}}$ values listed in Table $\ref{tab:SiPM_prope}$, $\it{P}_{CN}$ in the absence of an external field is found to be $\sim$ \SI{9}{\percent}, \SI{49}{\percent} and \SI{2.2}{\percent} for the FBK-RGB, FBK-LF and Hamamatsu VUV3 SiPMs, respectively. Figure~$\ref{fig:P_CN}$ shows $\eta_{P_{CN}}$ as a function of the external electric field strength. We observe a constant response with a maximum deviation of $\sim$~\SI{8}{\percent} compared to the value in the absence of any external electric field, for all three SiPMs tested.

\subsection{Results for Data Collected with Light}
\label{sec:PDE}
In this data set the SiPMs are illuminated by a blue LED, as described in section~\ref{sec:DAQ}. The LED signals are sent simultaneously to the SiPMs and a monitor PMT. The ratio, $\it{\eta_{PDE}}$, of the total SiPM output charge at each HV value, $\it{Q_{total}}$($\it{E}$), to that at \SI{0}{\kilo\volt} value, $\it{Q_{total}}$($\it{E}$ = 0), is computed as a function of the electric field strength at the SiPM surface:

\begin{equation}
\eta_{PDE} = \frac{Q_{total}(E)}{Q_{total}(E=0)}
\end{equation}

\begin{figure}
   \centering
   \includegraphics[width=0.9\linewidth]{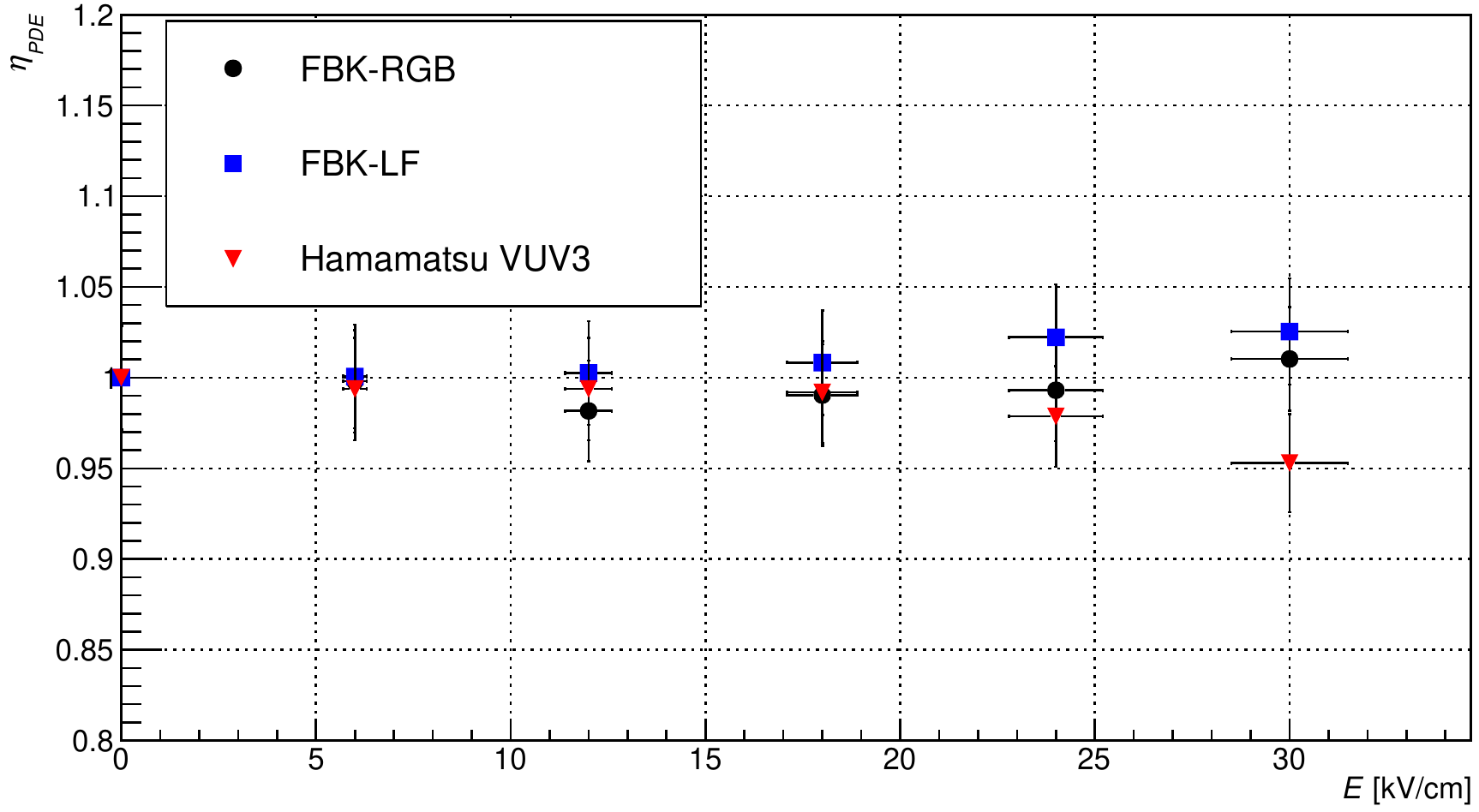}
   \caption{Dependence of the relative charge collected by the SiPMs under \SI{465}{\nano\meter} light illumination as a function of external electric field strength for FBK-RGB (black circles), FBK-LF (blue squares) and Hamamatsu VUV3 (red triangles).}
   \label{fig:relative_gain_light}
\end{figure}

Because the measurements in the dark show that the SiPM gains do not change with the external electric field, we can consider this measurement under LED illumination as a test of the stability of the SiPM photon detection efficiency ($\it{PDE}$) for \SI{465}{\nano\meter} light. The uncertainty in the relative $\it{PDE}$ measurement is dominated by the systematics in the instability of the monitor system as discussed in section ~$\ref{sec:DAQ}$. Figure~$\ref{fig:relative_gain_light}$ shows $\it{\eta_{PDE}}$ as a function of the external electric field values for the FBK-RGB, FBK-LF, and Hamamatsu VUV3 SiPMs. It can be seen that the SiPMs $\it{PDE}$ for \SI{465}{\nano\meter} light does not change with external electric fields, within \SI{5}{\percent} deviation compared to that at the absence of the external field. 

\subsection{I-V Curve Studies}
\label{sec:IV_Plots}

Because SiPM I-V curves can reveal subtle changes in their characteristics, we measured the I-V curves of each device at different external electric field strength as a cross-check. In this measurement, we connected the anode and cathode of each SiPM to a picoammeter (Keithley 6487~\cite{picoammeter}) at \SI{149}{\kelvin} and measured its leakage currents as a function of the bias voltage.  The bias voltage was incremented in steps of \SI{0.5}{\volt} up to $8-10$ V below the breakdown voltage (determined at room temperature), after which the step size was reduced to \SI{0.1}{\volt} to improve the accuracy of the breakdown voltage determination. This study was repeated twice: in the dark and with the LED light source operating in a continuous mode. The effective resolution of the system is dominated by noise pickup, which is on the order of \SI{100}{\pico\ampere}.  

\begin{figure}
   \centering
   \includegraphics[width=0.49\linewidth]{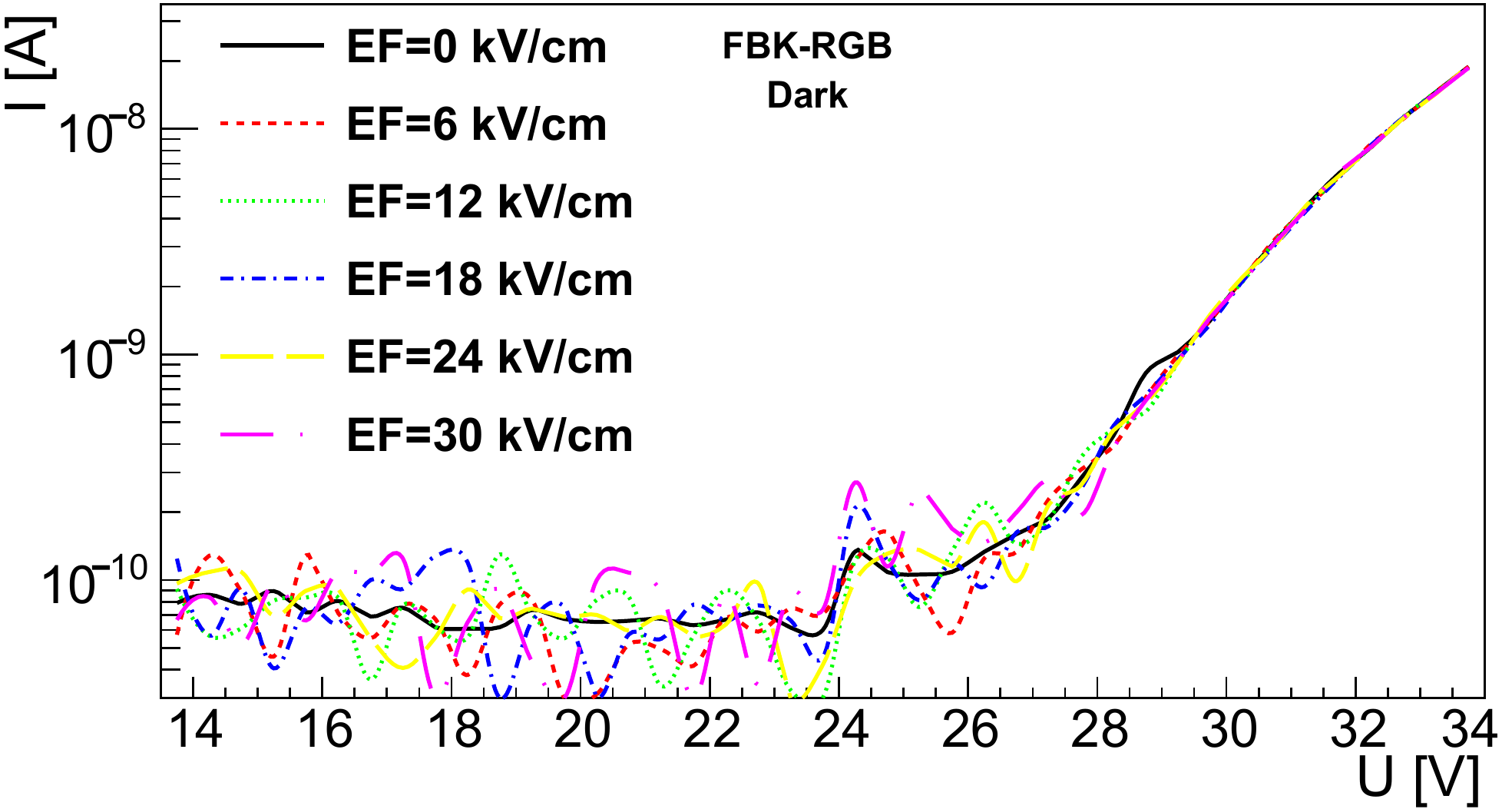} ~~\includegraphics[width=0.49\linewidth]{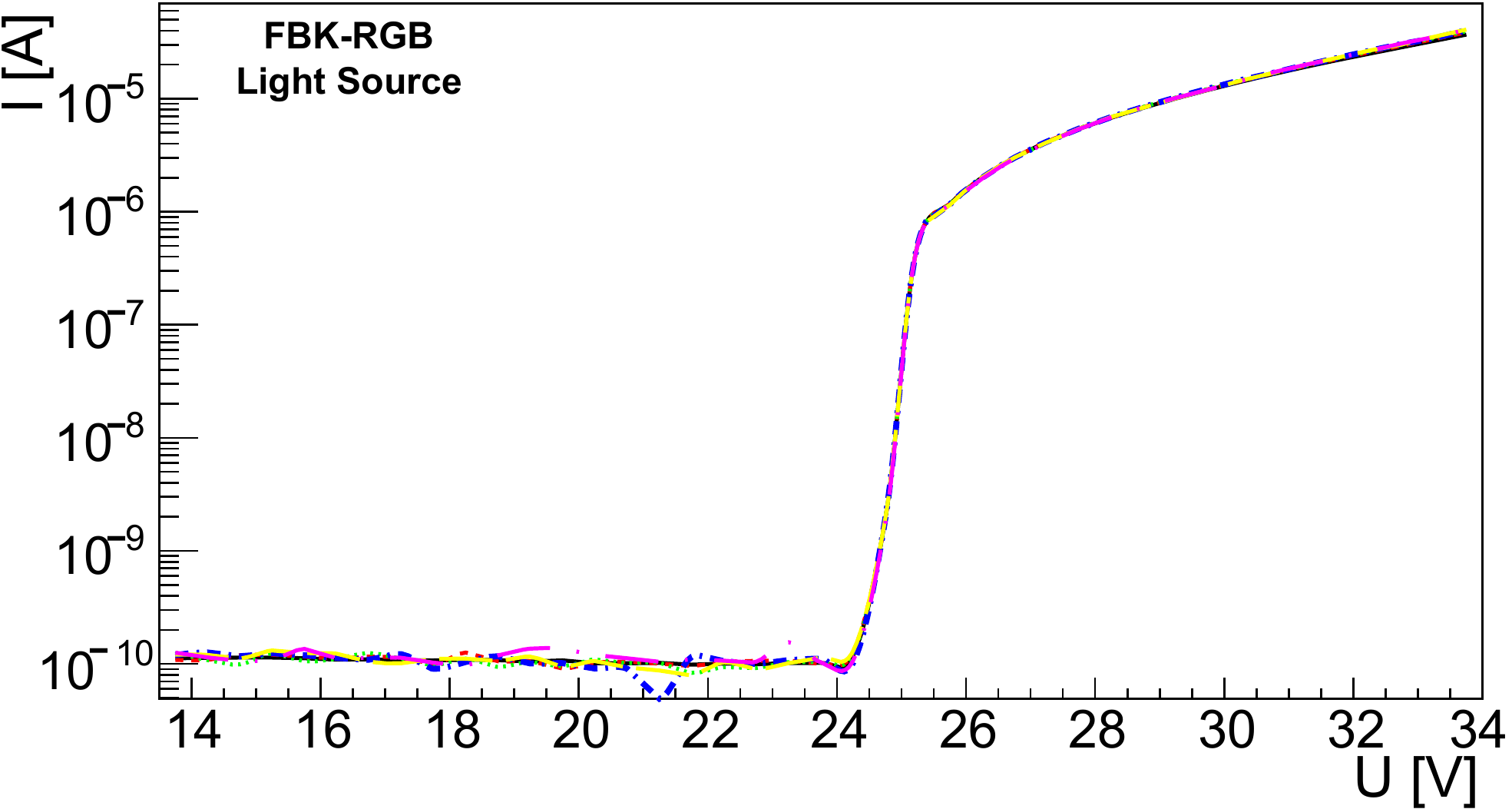}
   
   \vspace{2.5mm}
   
   \includegraphics[width=0.49\linewidth]{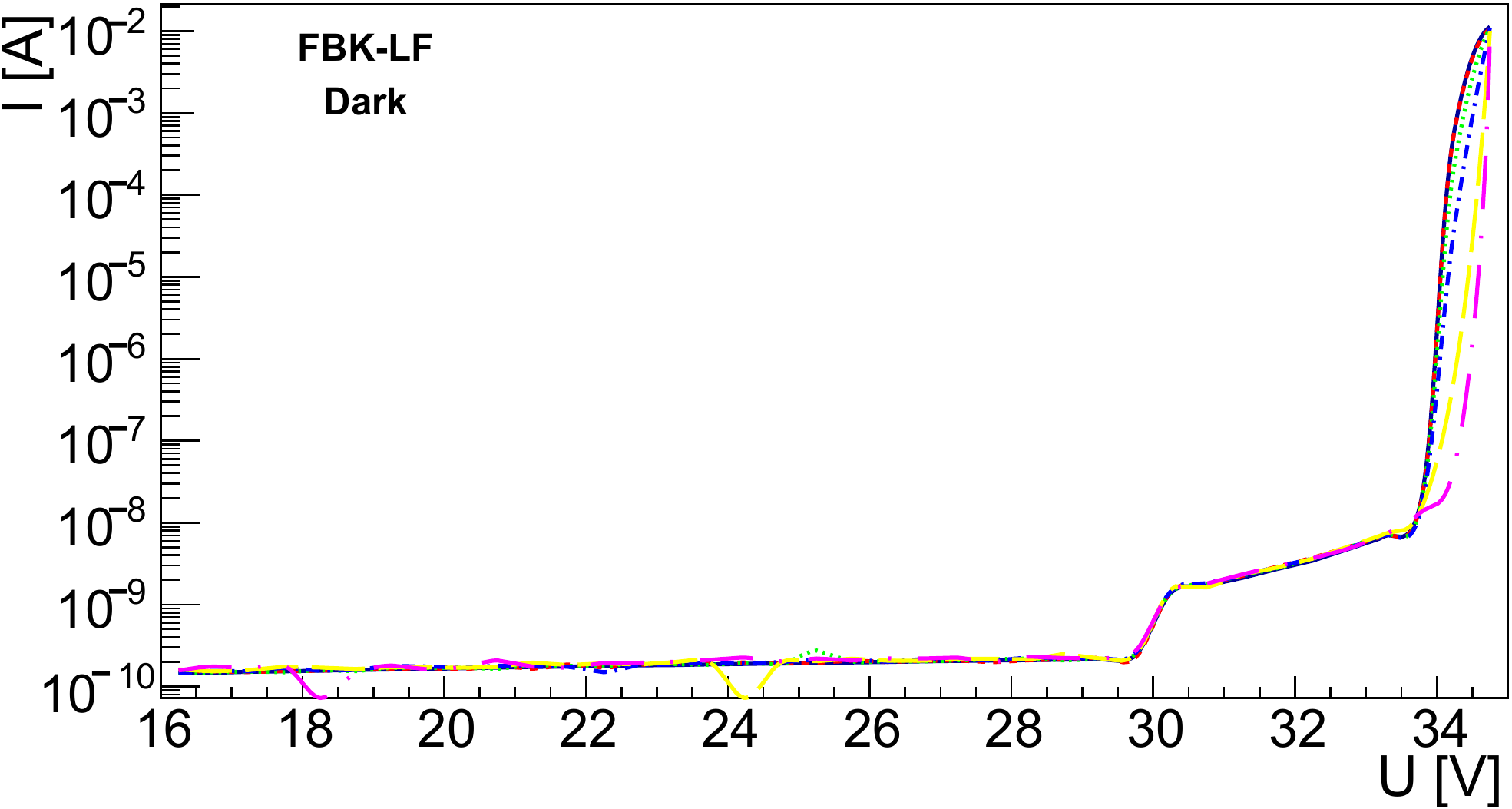} ~~\includegraphics[width=0.49\linewidth]{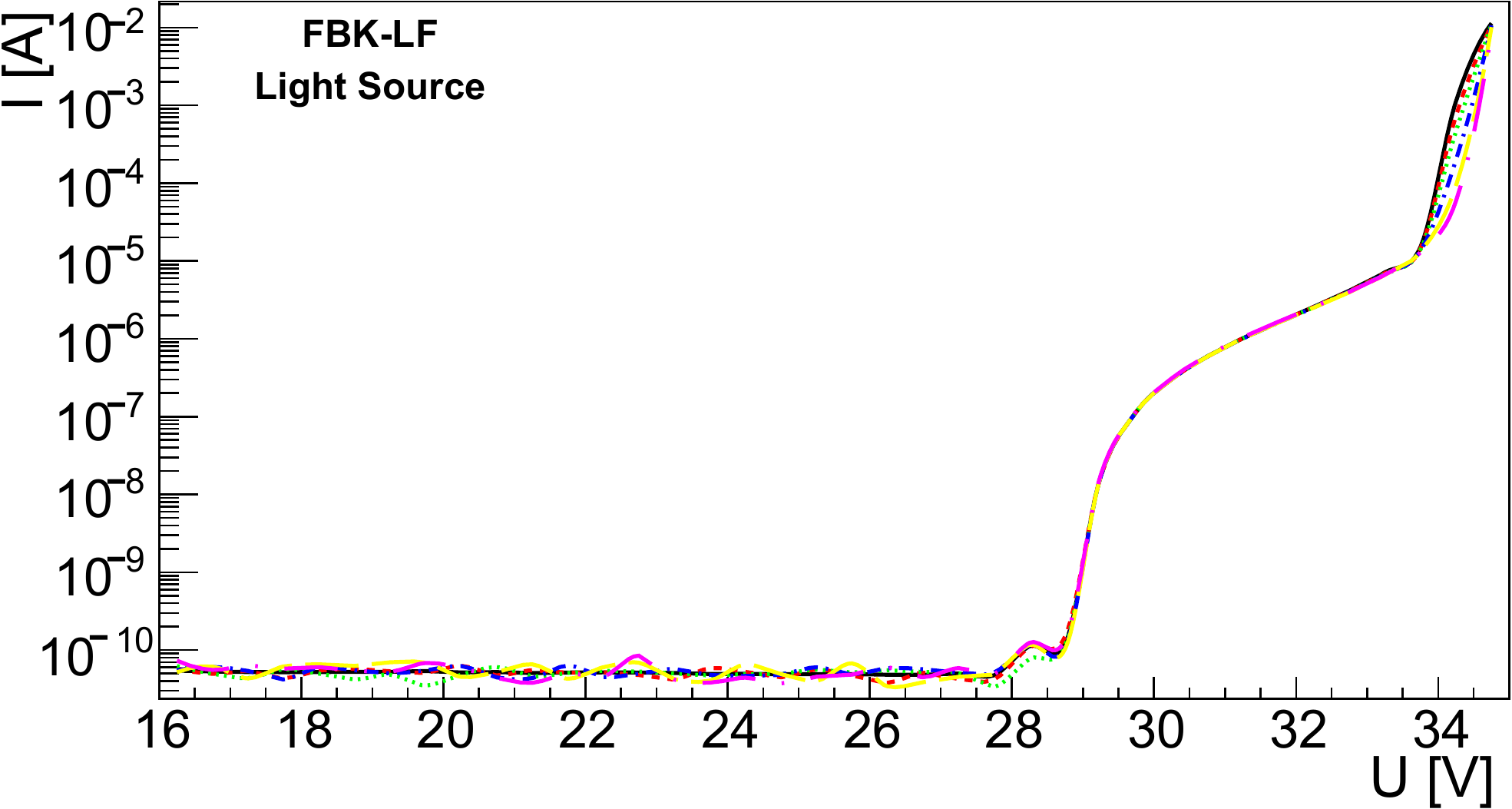}
   
   \vspace{2.5mm}
   
   \includegraphics[width=0.49\linewidth]{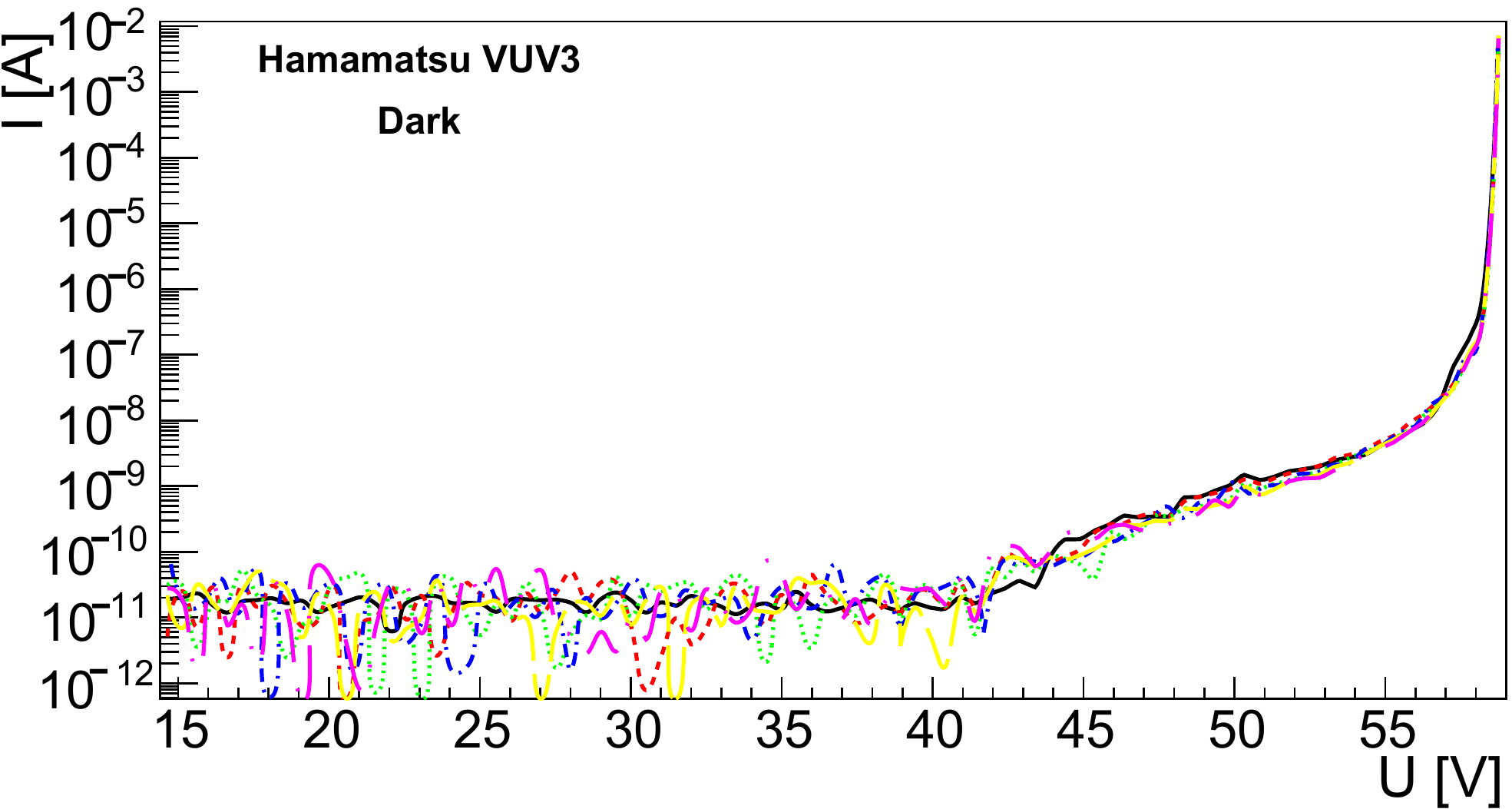} ~~\includegraphics[width=0.49\linewidth]{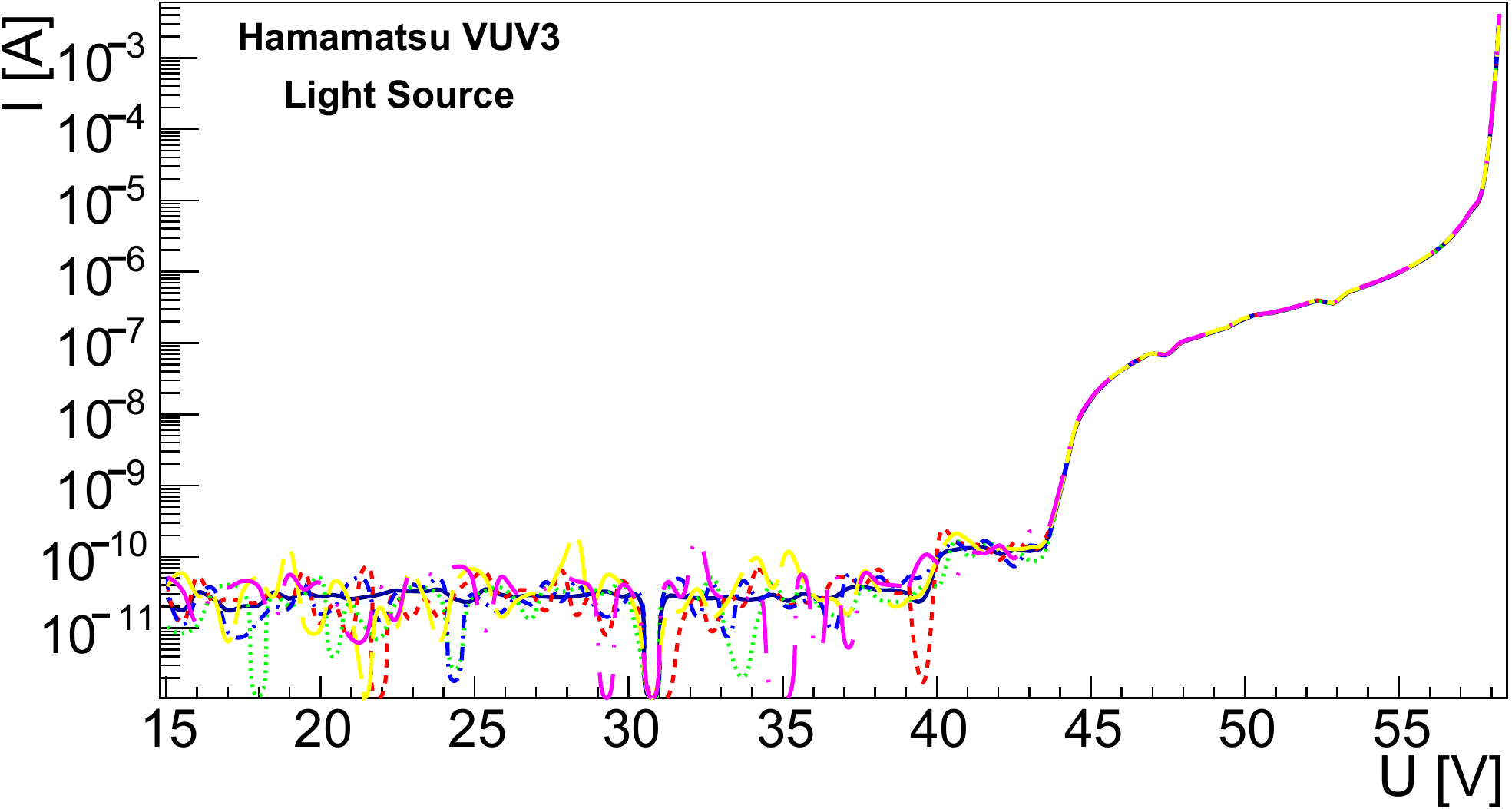}
   \caption{I-V curves for FBK-RGB (top), FBK-LF (middle) and Hamamatsu VUV3 (bottom) SiPMs. Data in the left column are taken in the dark, while data in the right column are taken under continuous illumination with a \SI{465}{\nano\meter} light from a blue LED.}
   \label{fig:IV_Curves}
\end{figure}

Figure~$\ref{fig:IV_Curves}$ (left column) shows the I-V curves for the three types of SiPMs measured in the dark, while the right column shows the results when the SiPMs are illuminated by the blue LED. The onset of breakdown is clearly visible with LED illumination at about \SI{24}{\volt}, \SI{29}{\volt} and \SI{44}{\volt} for the FBK-RGB, FBK-LF, and Hamamatsu VUV3 SiPMs, respectively. The onset of breakdown is less obvious without illumination because the dark current is very low at \SI{149}{\kelvin}. Within the resolution of the measurement the onset of breakdown does not change with external electric field.  The FBK-LF and Hamamatsu VUV3 SiPMs also clearly show a runaway transition, at about \SI{34}{\volt} and \SI{57}{\volt} respectively. Such a transition is electric field independent for the Hamamatsu device, while it shows a slight dependence on the electric field for the FBK-LF device. The runaway transition occurs when the correlated avalanche rate approaches unity, i.e. when the avalanche production becomes self-sustaining. The operating point of SiPMs is in the span in between the breakdown and runaway voltages. In this condition the mean current scales linearly with the rate of avalanches generated thermally or by photons.

\section{SiPMs Visual Inspection}
\label{sec:vis_ins}
A visual inspection of the SiPM devices was carried out at the end of the tests using an optical microscope, the Rational VMS-1510F system~\cite{microscope}, with magnification power ranging from 20x to 128x. The surfaces of the devices were carefully inspected, and some photographs of specific locations were taken, before and after the high voltage tests. No visible evidence for damage was found on the outer surface of the SiPMs or at the microcell level. 

\section{Conclusions}
\label{sec:conclusions}

We investigated the effect of external electric fields on the operation of several SiPM devices in liquid CF$_{4}$ at \SI{149}{\kelvin}.  Our experiments show that the performance of these SiPMs, Hamamastu VUV3,  FBK-RGB, and FBK-LF, is not affected significantly by external electric fields perpendicular to the surface, up to \SI{30}{\kilo\volt\per\cm}. We measured the I-V curves of the devices, both in the dark and under blue light illumination, confirming that the basic operation parameters of the devices do not change with external electric field. The SiPM devices were also inspected under a microscope and no visible damage was observed.  

In summary, our study demonstrates that SiPMs can operate normally in high electric fields at cryogenic temperatures, which bodes well for their use in experiments such as nEXO. In the future we plan to study the long term stability of SiPMs operating in high external electric fields and possible effects due to surface charge build-up. 

\acknowledgments{
This work has been supported by CAS and ISTCP in China, DOE and NSF in the United States, NSERC, CFI, FRQNT and NRC in Canada, IBS in South Korea and RFBR in Russia. X.L.~Sun was supported, in part, by the Young Scientists Fund of the Chinese National Natural Science Foundation.
}

\bibliographystyle{JHEP} 


\bibliography{references}

\providecommand{\href}[2]{#2}\begingroup\raggedright\begin{thebibliography}{10}

\bibitem{BONDARENKO2000}
G.~Bondarenko et~al., \emph{\it{Limited Geiger-Mode Microcell Silicon
  Photodiode: New Results}},
  \href{http://dx.doi.org/https://doi.org/10.1016/S0168-9002(99)01219-X}{\emph{NIM}
  {\bfseries A442} (2000) 187--192}.

\bibitem{nEXOSiPMpaper}
{\scshape \it{n}EXO} collaboration, A.~Jamil et~al., \emph{\it{VUV-Sensitive
  Silicon Photomultipliers for Xenon Scintillation Light Detection in nEXO}},
  \href{https://arxiv.org/abs/\tt arxiv:1806.02220
  [physics.ins-det]}{{\ttfamily \tt arxiv:1806.02220 [physics.ins-det]}}.

\bibitem{Ostrovskiy-SiPM2015}
I.~Ostrovskiy et~al., \emph{\it{Characterization Of Silicon Photomultipliers
  for nEXO}}, \href{http://dx.doi.org/10.1109/TNS.2015.2453932}{\emph{IEEE
  Trans. Nucl. Sci.} {\bfseries 62} (Aug, 2015) 1825--1836}.

\bibitem{PhysRevC.97.065503}
{\scshape \it{n}EXO} collaboration, J.~B. Albert et~al., \emph{\it{Sensitivity
  and Discovery Potential of the Proposed nEXO Experiment to Neutrinoless
  Double-$\ensuremath{\beta}$ Decay}},
  \href{http://dx.doi.org/10.1103/PhysRevC.97.065503}{\emph{Phys. Rev. C}
  {\bfseries 97} (2018) 065503}.

\bibitem{nEXOpCDR}
{\scshape \it{n}EXO} collaboration, S.~Al~Kharusi et~al., \emph{\it{nEXO
  \it{Pre-Conceptual Design Report}}},  \href{https://arxiv.org/abs/\tt
  arxiv:1805.11142 [physics.ins-det]}{{\ttfamily \tt arxiv:1805.11142
  [physics.ins-det]}}.

\bibitem{FUJII2015293}
K.~Fujii et~al., \emph{\it{High-Accuracy Measurement of the Emission Spectrum
  of Liquid Xenon in the Vacuum Ultraviolet Region}},
  \href{http://dx.doi.org/https://doi.org/10.1016/j.nima.2015.05.065}{\emph{NIM}
  {\bfseries A795} (2015) 293 -- 297}.

\bibitem{comsol}
\normalfont{COMSOL Multiphysics}. \url{https://www.comsol.com}.

\bibitem{SiPM_MagF}
R.~Hawkes et~al., \emph{\it{Silicon photomultiplier performance tests in
  magnetic resonance pulsed fields}},
  \href{http://dx.doi.org/10.1109/NSSMIC.2007.4436860}{\emph{2007 IEEE Nuclear
  Science Symposium Conference Record} {\bfseries 5} (Oct, 2007) 3400--3403}.

\bibitem{SPAD_breakdown}
W.~G. Oldham, R.~R. Samuelson and P.~Antognetti, \emph{\it{Triggering phenomena
  in avalanche diodes}}, {\emph{IEEE Trans. Elect. Dev.} {\bfseries 19} (1972)
  1056--1060}.

\bibitem{MOS-FET}
S.~M. Sze and K.~K. Ng, \emph{\it{Physics of semiconductor devices}}.
\newblock John Wiley \& Sons, 2006.

\bibitem{Anderson:2017zun}
T.~Anderson et~al., \emph{\it{Evaluation of Silicon Photomultipliers for use as
  Photosensors in Liquid Xenon Detectors}},  \href{https://arxiv.org/abs/\tt
  arxiv:1706.05371 [physics.ins-det]}{{\ttfamily \tt arxiv:1706.05371
  [physics.ins-det]}}.

\bibitem{HVmodule}
\normalfont{Spellmanhv}. \url{http://www.spellmanhv.com}.

\bibitem{CF4}
N.~Gee and G.~R. Freeman, \emph{\it{Relative permittivities of 10 organic
  liquids as functions of temperature}},
  \href{http://dx.doi.org/10.1006/jcht.1993.1163}{\emph{J. Chem. Thermodyn.}
  {\bfseries 25} (April, 1993) 549--554}.

\bibitem{Amey:1964}
R.~L. Amey and R.~H. Cole, \emph{\it{Dielectric Constants of Liquefied Noble
  Gases and Methane}}, \href{http://dx.doi.org/10.1063/1.1724850}{\emph{J.
  Chem. Phys.} {\bfseries 40} (1964) 146--148}.

\bibitem{CF4-mp}
W.~M. Haynes, \emph{\it{CRC Handbook of Chemistry and Physics}}.
\newblock CRC Press LLC, Boca Raton, 95th~ed., 2014-2015.

\bibitem{fbk}
\normalfont{Fondazione Bruno Kessler}. \url{https://www.fbk.eu}.

\bibitem{hamamatsu}
\normalfont{Hamamatsu}. \url{http://www.hamamatsu.com}.

\bibitem{Serra2013}
N.~Serra et~al., \emph{\it{Characterization of New FBK SiPM Technology for
  Visible Light Detection}},
  \href{http://dx.doi.org/10.1088/1748-0221/8/03/P03019}{\emph{JINST}
  {\bfseries 8} (2013) P03019}.

\bibitem{preamp}
\normalfont{Photonique SA}, ``\textit{AMP-0604 and AMP-0611 Amplifiers Data
  Sheet}.'' \url{http://www.photonique.ch/}.

\bibitem{caen}
\normalfont{CAEN}, ``\textit{DT5751 Digitizer Data Sheet}.''
  \url{http://www.caen.it}.

\bibitem{labview}
\normalfont{National Instruments}, ``Labview.''
  \url{http://www.ni.com/labview}.

\bibitem{ROOT}
R.~Brun and F.~Rademakers, \emph{\it{ROOT - An object oriented data analysis
  framework}},
  \href{http://dx.doi.org/https://doi.org/10.1016/S0168-9002(97)00048-X}{\emph{NIM}
  {\bfseries A389} (1997) 81 -- 86}.

\bibitem{PMT}
\normalfont{ET~Enterprises}.
  \url{https://my.et-enterprises.com/search?search-source=PMT&q=9364U}.

\bibitem{AdvanSiD}
\normalfont{AdvanSiD}, ``\textit{Application note: Introduction to SiPMs}.''
  \url{http://advansid.com}.

\bibitem{SensL}
\normalfont{SensL}, ``\textit{Technical Note: Introduction to SiPM}.''
  \url{http://www.sensl.com}.

\bibitem{picoammeter}
\normalfont{Keithley}. \url{http://www.keithley.com}.

\bibitem{microscope}
\normalfont{Rational Precision Instrument Co. LtD}.
  \url{http://www.rational-en.com/Productshow-71-443.html}.

\end{thebibliography}\endgroup

\end{document}